\documentclass[aps,prd,twocolumn,nofootinbib,floatfix]{revtex4-1}
\usepackage{setspace}
\usepackage{amsthm}
\theoremstyle{definition}

\usepackage{url}
\usepackage{graphicx}
\usepackage{dcolumn}
\usepackage{multirow}
\usepackage{subfigure}

\usepackage{times,mathptm}
\usepackage{float}
\usepackage{color}
\usepackage{amsmath,amsfonts}
\usepackage{mathptmx}
\usepackage{mathrsfs}
\usepackage{bbm}
\usepackage{bm}
\usepackage{xfrac}

\newcommand{\beq}{\begin{equation}}
\newcommand{\eeq}{\end{equation}} 
\newcommand{\bea}{\begin{eqnarray}}
\newcommand{\eea}{\end{eqnarray}}

\newcommand{\ua}{\uparrow}
\newcommand{\da}{\downarrow}

\newcommand{\E}{\mathcal{E}}

\newcommand{\tU}{\widetilde{U}}

\newcommand{\C}{\mathcal {C}}

\renewcommand{\d}{\delta}

\renewcommand{\l}{\lambda}

\newcommand{\Dbar}{\overline{\Delta}}

\newcommand{\tc}{\widetilde{c}}

\renewcommand{\l}{\lambda}
\renewcommand{\b}{\beta}
\renewcommand{\a}{\alpha}

\newcommand{\tr}{\text{Tr}}

\renewcommand{\o}{\omega}
\newcommand{\bx}{\mathbf{x}}
\newcommand{\by}{\mathbf{y}}

\newcommand{\vx}{{\vec{x}}}
\newcommand{\vy}{{\vec{y}}}
\newcommand{\vz}{\vec{z}}
\newcommand{\vk}{{\vec{k}}} 
\newcommand{\vq}{{\vec{q}}}

\newcommand{\n}{\nu}
\newcommand{\m}{\mu}

\newcommand{\g}{\gamma}

\newcommand{\e}{\epsilon}
\newcommand{\s}{\sigma}
\renewcommand{\k}{\kappa}
\newcommand{\D}{\Delta}

\newcommand{\vn}{\vec{n}}

\newcommand{\vA}{\vec{A}}

\newcommand{\vp}{\vec{p}}
\renewcommand{\vn}{\vec{n}}

\newcommand{\N}{{\cal N}}

\newcommand{\Hp}{H_I^{(1)}}
\newcommand{\Hd}{H_I^{(2)}}
\renewcommand{\th}{\theta}

\newcommand{\oh}{\frac{1}{2}}

\newcommand{\dg}{\dagger}
\newcommand{\non}{\nonumber}

\newcommand{\rf}[1]{(\ref{#1})}
\newcommand{\ra}{\rightarrow}
\newcommand{\pa}{\partial}
\renewcommand{\vec}[1]{\bm #1}

\usepackage{ulem}

\begin{document}

\title{The nature of symmetry breaking in the superconducting ground state} 

\bigskip
\bigskip

\author{Kazue Matsuyama and Jeff Greensite}
\affiliation{Physics and Astronomy Department \\ San Francisco State
University   \\ San Francisco, CA~94132, USA}
\bigskip
\date{\today}
\vspace{60pt}
\begin{abstract}

 
   The order parameters which are thought to detect U(1) gauge symmetry breaking in a superconductor are
both non-local and gauge dependent. For that reason they are also ambiguous as a guide to phase structure.  We point out that a global subgroup of the local U(1) gauge symmetry may be regarded, in analogy to non-abelian theories, as a ``custodial'' symmetry affecting the matter field alone, and construct, along the lines of our previous work, a new gauge-invariant criterion for breaking symmetries of this kind.  It is shown that spontaneous breaking of custodial symmetry
is a necessary condition for the existence of spontaneous symmetry breaking of a global subgroup of the (abelian or non-abelian) gauge group in any given gauge, and a sufficient condition for the existence of spontaneous breaking of a global subgroup of the gauge group
in some gauge.  As an illustration we compute numerically,  in the lattice version of the Ginzburg-Landau model, the phase boundaries of the theory and the order parameters associated with various symmetries in each phase.  


\end{abstract}

\maketitle

\singlespacing
   
\section{\label{intro} Introduction}   
   
    Superconductivity is the simplest example of a so-called dynamically broken gauge symmetry.
In view of the Elitzur theorem, which states that gauge symmetry is unbreakable either dynamically or spontaneously,
this characterization deserves closer scrutiny.  What symmetry, exactly, is broken?  And in which operators is that
breaking manifest?  The issue is largely a conceptual one, since the BCS theory seems perfectly adequate for conventional (non cuprate) superconductors, but these questions seem relevant not just to a deeper understanding of superconductivity, but also to a better understanding of any theory which is claimed to break a gauge symmetry, whether spontaneously or dynamically.
   
   Certainly the ground state of a superconductor, and in fact any physical state, must be invariant under infinitesimal and, more generally, local gauge transformations of the dynamical fields; this is required by the Gauss Law condition, and the vanishing of locally non-invariant operators in the ground state is guaranteed by the Elitzur theorem \cite{Elitzur:1975im}.  But neither Gauss's Law nor the Elitzur theorem forbids the breaking of a global symmetry, and in fact there is a global U(1) subgroup of the gauge symmetry which appears to be broken by the superconducting ground state.    But the order parameter which has been proposed to detect the breaking of this gauge symmetry is itself gauge dependent, and the magnitude of the order parameter, including whether it is zero or non-zero, depends on the gauge choice, as shown below in an effective model.  In view of this fact, is it possible to construct a gauge invariant criterion which distinguishes the symmetric phase from the symmetry broken phase in U(1) gauge theories, and in gauge-Higgs theories in general?  That is the question we would like to address here.

    To fix notation, let $c_\s(x), c^\dg_\s(x)$ denote the electron operators with spin index $\s$, 
transforming as
\bea
          c_\s(x) &\ra& e^{i \th(x)} c_\s(x) ~~,~~ c^\dg_\s(x) \ra e^{-i \th(x)} c^\dg_\s(x) \non \\
           A_\m(x) &\ra& A_\m(x) + {1\over e} \pa_\m \th(x)
\label{gtrans}
\eea
under a local gauge transformation.   A global U(1) subgroup of the gauge group is defined by the set of transformations with $\th(x)=\th$ independent of space.  But this can be regarded as a global symmetry pertaining to the matter sector of the theory alone,  since the $A_\m$ gauge field is unaffected by such transformations.  For this reason, adopting a term from the electroweak sector of the Standard Model, it may be regarded as a type of ``custodial symmetry."  Of course, the name we choose to assign to a symmetry may be just a matter of words (although we will support our preference in section \ref{sec5}), but the choice of order parameter to detect symmetry breaking is not just semantic.  In the context of superconductivity it is usually the expectation value of the Cooper pair creation operator which is said to detect the breaking of gauge invariance in the BCS ground state.  But since this operator transforms under local as well as global transformations, it can only serve as an order parameter for global symmetry breaking in a fixed gauge.  The spontaneous breaking of a ``remnant''
gauge symmetry, i.e.\ a global symmetry which remains after gauge fixing, is known to be ambiguous, in that  the symmetry breaking transition depends on the gauge choice \cite{Caudy:2007sf}.  This ambiguity is consistent with the theorem proved by Osterwalder and Seiler \cite{Osterwalder:1977pc}, whose consequences were elaborated by Fradkin and Shenker \cite{Fradkin:1978dv}.   There is, moreover, the issue of the Goldstone theorem: if a global symmetry (whether or not we call it a gauge symmetry) breaks spontaneously, then there ought to exist gapless excitations, which are not found in superconductors.  

    This raises the question of whether we can find a gauge-invariant criterion for the breaking of a global symmetry characterized by $\th(x)=\th$ and, if so, how the Goldstone theorem is evaded.  We will address this
question, along the lines of our recent work \cite{Greensite:2018mhh} in non-abelian gauge-Higgs theories, in the context of a lattice version of Ginzburg-Landau theory.  But first, in section \ref{sec2}, we present a generalized version of the usual BCS ground state, this time incorporating quantized gauge field and ion degrees of freedom, which satisfies the physical state constraint (i.e.\ Gauss's Law), and which illustrates the gauge dependent nature of the usual order parameter for symmetry breaking.   In passing we derive, in section \ref{sec3}, the momentum-dependent mass function of a transverse photon from minimization of the ground state energy of the generalized BCS state.  In section \ref{sec4} we show explicitly, in the context of an effective Ginzburg-Landau theory, the ambiguity of spontaneous gauge symmetry breaking, in the sense that the location (and even the existence) of a symmetry breaking transition of this kind actually depends on the gauge choice.  

    The main point of this paper is presented in Section \ref{sec5}.  In that section we
introduce a gauge-invariant criterion for custodial symmetry breaking in the effective Ginzburg-Landau theory, comparing the thermodynamic and symmetry-breaking transition lines obtained via lattice Monte Carlo simulations, and
show that custodial symmetry breaking is a necessary condition for the spontaneous breaking of a global subgroup of the gauge symmetry in any given gauge.   Here we also make contact with related observations in non-abelian gauge Higgs theory, and we propose that the Higgs phase should be {\it defined}, in a gauge-invariant manner, as the phase of broken custodial symmetry.  Our conclusions are in section \ref{conclude}.

\section{\label{sec2} A physical version of the BCS ground state}

    In a Hamiltonian formulation of gauge theory in a physical gauge,  Gauss's Law is implemented either as a constraint on physical states, as in temporal gauge, or by explicitly solving for the longitudinal electric field in terms of the other degrees
of freedom, as in Coulomb gauge, which introduces long-range interaction terms in the Hamiltonian.  In this article we opt for temporal $A_0=0$ gauge, since the issues we wish to address are clearest in that gauge.  The Gauss law constraint, which is
\beq
       (\nabla \cdot E - \rho) ~  \Psi = 0
\label{physical}
\eeq
for all physical states, is equivalent to the requirement that $\Psi$ is invariant with
respect to infinitesimal gauge transformations.\footnote{This is in the absence of external, non-dynamical charges.  If such
charges are present in the system, then the wavefunctional transforms covariantly at the locations of those charges.}   In the context of the BCS theory, however, there is a technical
issue regarding indefinite particle number which must first be addressed.

    In the standard BCS treatment one considers a region of finite volume with periodic boundary conditions, and
the BCS ground state is not an eigenstate of particle number.  If the number of positively charged ions is fixed, then
the BCS ground state is also not an eigenstate of electric charge.  There is then a difficulty in applying Gauss's law,
because this law cannot be satisfied in a volume with periodic boundary conditions if the net charge is non-zero.
Formally, Poisson's equation does not have a solution in a volume with periodic boundary conditions unless the
net charge vanishes. One option is insert a constraint which correlates electron and ion number.  A technically simpler
alternative is to embed the finite volume solid in an infinite volume space, and allow the Coulomb electric field to escape from the solid into the surrounding space.  So let us begin
with the electromagnetic Hamiltonian (in Heaviside-Lorenz units)
\bea
          H_A &=& \oh \int d^3 x ({\bf E}^2 + {\bf B}^2) \ ,
\eea
with
\beq   
            [A_i(x),E_j(y)] = i \d_{ij} \d^3(x-y)
\eeq
in temporal gauge.  Decomposing the gauge field into longitudinal and transverse components,
\bea
          A_i(x) &=& \int {d^3k \over (2\pi)^{3/2} } A_i(k) e^{i\vk \cdot \vx} \non \\
          A_i(k) &=& {ik_i\over k} \a(k) + \sum_\l \e_i(k,\l) A(k,\l) \ ,
\eea
where $\e_i(k,\l)$ is the usual transverse polarization vector, we find
\beq
           H_A = \oh \int d^3x \Bigl\{ E_L^2(x) +  \sum_\l  (E_T^2(x,\l) + A(x,\l)(-\nabla^2)A(x,\l)) \Bigr\}   \ ,
\eeq
where
\bea
          [\a(x),E_L(y)] &=& i \d^3(x-y) \non \\
          \left[ A(x,\l),E_T(y,\l') \right] &=& i \d_{\l \l'} \d^3(x-y) \ .
\eea

    If we are only interested in quantum fluctuations of the $A$ field deep inside the solid, then it
is sufficient to neglect the fluctuations of the transverse $A$ field outside the solid, and consider only
\bea
           H_A & & = \oh \int d^3x \Bigl\{ E_L^2(x) \non \\
           & & + \oh \int_V d^3x \sum_\l  (E_T^2(x,\l) + A(x,\l)(-\nabla^2)A(x,\l)) \Bigr\}   \ ,
\eea
where the second spatial integration is restricted to the volume $V$ of the solid.  We may even impose periodic
boundary conditions on $A(x,\l),E(x,\l)$ in this region, on the grounds that the obvious errors which are thereby 
introduced at the boundaries are unimportant in the thermodynamic limit.   However, for reasons already stated,
we cannot impose such boundary conditions on the longitudinal degrees of freedom described by $\a(x), E_L(x)$.
If there is any net charge in the solid, then the corresponding Coulomb electric field necessarily extends outside
the solid, so the integration region for these degrees of freedom must be over all space.

    We then consider the total Hamiltonian for quantized electron and electromagnetic degrees of freedom
\beq
         H =  H_{BCS} + H_A \ ,
\label{H}
\eeq
where
\bea
           H_{BCS} &=& \int d^3 x ~ c^\dg_\s(x) \left[ {1\over 2m}(-i \nabla -e \vA)^2 - \e_F \right] c_\s(x) \non \\
           &-& {g\over V}  {\sum_k}'   {\sum_{k'}}' c^\dg_\ua(k) c^\dg_\da(-k) c_\da(k') c_\ua(-k') \ . \non \\
 \label{HBCS0}
 \eea
 Here we have defined
 \bea
 {\sum_k}' [\cdot \cdot \cdot]  &\equiv& \sum_k \th(\o_D - |\e_k|) [\cdot \cdot \cdot]  \ ,
 \eea
 with
 \bea
 \e_k &=& {k^2 \over 2m} - \e_F \ ,
 \eea
and $\o_D,\e_F$ are the Debye frequency and Fermi energy respectively.  A finite volume $V$ with periodic boundary conditions is assumed. We have the usual anticommutation relation among fermion operators
 \beq
   \{c_\s(x),c^\dg_{\s'}(y)\} = \d_{\s \s'} \d^3(x-y)  \ ,
\eeq
 and $\s= \ua, \da$ denotes spin up/down.  We treat the ions as static sources of charge $+pe$ located at fixed points
 $\vx_n, n=1,2,...,N_{ions}$.  
 Our goal is to find an approximate ground state of the Hamiltonian in \rf{H}, which
 obeys the physical state condition \rf{physical} with charge density operator
 \beq
           \rho(x) =  -e c^\dg_\s(x) c_\s(x) + pe \sum_{\vn} \d^3(x - x_n)  \ .
\label{chargedensity}
 \eeq
The approximations involve the usual BCS mean field approach, and a neglect
of correlations between ions and electrons in the ground state.  It is assumed that the effect of such correlations has already been accounted for in the attractive four fermi interaction in $H_{BCS}$.
 
    The physical state condition cannot be satisfied by the ground state of $H$ as it stands, because the four-fermi term in
$H_{BCS}$ is completely gauge non-invariant.  There is, however, a simple fix for that, at least 
if we suppose that $H_{BCS}$
is the correct expression in Coulomb gauge.  Let us introduce the phase factor
\bea
            e^{i\g_x} &=& \exp\left[i {e\over 4\pi} \int d^3z ~ A_i(z) {\pa \over \pa z_i} {1\over |\vx -\vz|} \right] \non \\
            &=& \exp\left[i e \int {d^3k \over (2\pi)^{3/2}} {\a(k) \over k} e^{i\vk \cdot \vx} \right]  \ ,
\eea
Note that the $z$-integration is over all space, and not just within the solid.
Under a local gauge transformation \rf{gtrans}, 
\beq
            e^{i\g_x} \ra e^{i \th(x)} e^{i\g_x} \ .
\eeq
Next introduce operators which are invariant under such local transformations
\beq
           \tc_{\s}(x) = c_{\s}(x) e^{-i\g_x} ~~,~~ \tc^\dg_{\s}(x) = c^\dg_{\s}(x) e^{i\g_x} \ ,
\eeq
with $\tc_\s(k),\tc^\dg(k)$ the corresponding Fourier transforms.  The commutator relations for $\tc, \tc^\dg$ are identical to those for $c, c^\dg$.  We therefore simply replace the $c,c^\dg$
operators in the four-fermi term of the ``tentative'' $H_{BCS}$ by $\tc,\tc^\dg$.  Moreover we note the equality in the kinetic terms
\bea 
& &\int d^3 x ~ c^\dg_\s(x) \left[ {1\over 2m}(-i \nabla -e \vA)^2 - \e_F \right] c_\s(x) \non \\
& &  \qquad = \int d^3 x ~ \tc^\dg_\s(x) \left[ {1\over 2m}(-i \nabla -e \vA^T)^2 - \e_F \right] \tc_\s(x)   \ ,
\label{inspection}
\eea
where $\vA^T$ is the transverse, gauge invariant part of the $A$-field
\bea
           A^T_i(x) &=& \int {d^3k \over (2\pi)^{3/2}} A_i^T(k) e^{i\vk \cdot \vx} \non \\
           A^T_i(k) &=& \sum_\l \e_i(k,\l) A(k,\l) \ .
 \eea
This equality can be seen by inspection, just by noting that both sides of \rf{inspection} are gauge invariant, and are trivially equal to one another in Coulomb gauge, where $A=A^T$ and $\g_x=0$.  We then rewrite the BCS Hamiltonian in the gauge invariant form
\bea
           H_{BCS} &=& \int d^3 x ~ \tc^\dg_\s(x) \left[ {1\over 2m}(-i \nabla -e \vA^T)^2 - \e_F \right] \tc_\s(x) \non \\
           &-& {g\over V} {\sum}'_k {\sum}'_{k'} \tc^\dg_\ua(k) \tc^\dg_\da(-k) \tc_\da(k') \tc_\ua(-k') \ . 
 \label{HBCS1}
 \eea
 Although $H_{BCS}$ is entirely gauge invariant, $\tc$ is not quite invariant.  Let us define
 \beq
            \th(x) = \th_0 + \varphi(x)   ~~\mbox{where~~} \int d^3x \varphi(x) = 0 \ ,
\eeq
then $\tc, \tc^\dg$ transform under an arbitrary transformation \rf{gtrans} as
\beq 
             \tc(x) \ra e^{i\th_0} \tc(x) ~~~,~~~ \tc^\dg(x) \ra e^{-i\th_0} \tc^\dg(x) \, 
\label{global}
\eeq
and this is because the gauge field $A_k$, and in consequence $e^{i\g_x}$, are unchanged under the global
group of transformations, which we may or may not refer to as gauge transformations, defined by $\th(x)=\th_0$.

   Now repeating the usual steps of the mean field approach, we define
\bea
      \C^\dg &=& {\sum_k}'  \tc^\dg_\ua(k) \tc^\dg_\da(-k) ~~,~~ 
      \C = {\sum_k}'  \tc_\da(-k) \tc_\ua(k)  \non \\
            \D &=& -{g \over V} \langle \C \rangle_{BCS}  ~~,~~
      \Dbar = -{g \over V} \langle \C^\dg \rangle_{BCS} \non \\
      \d\C &=& \C - \langle \C \rangle_{BCS}  \ ,
\eea
where $\langle...\rangle_{BCS}$ indicates the expectation value in the BCS ground state.
Let us also define
\beq
          \D_k = \left\{ \begin{array}{cc}
                                \D & |\e_k| < \o_D \cr
                                0   & |\e_k|  \ge \o_D \end{array} \right. \ .
\eeq
Then
\bea
H_{BCS} &=& \int d^3 x ~ \tc^\dg_\s(x) \left[ {1\over 2m}(-i \nabla -e \vA^T)^2 - \e_F \right] \tc_\s(x) \non \\
& &  + \sum_k \left[ \Dbar_k \tc_\da(-k) \tc_\ua(k) + \D_k \tc^\dg_\ua(k) \tc^\dg_\da(k) \right] + {V\over g}\Dbar \D \non \\
& & + {g \over V}\d\C^\dg \d\C \ .
\eea
The mean field approximation drops the last term, and we have
\bea
         H^{mf}_{BCS} &=&  H_0 +   \Hp + \Hd \ ,
\eea
where
\bea
         H_0 &=&  \int d^3 x ~ \left\{ \tc^\dg_\s(x) \left[ {1\over 2m}(-\nabla^2) - \e_F \right] \tc_\s(x) \right\}  \non \\
                 & &  \qquad + \sum_k \left[ \Dbar_k \tc_\da(-k) \tc_\ua(k) + \D_k \tc^\dg_\ua(k) \tc^\dg_\da(k) \right] 
                 + {V\over g}\Dbar \D \non \\ 
         \Hp &=& -{ie \over m} \int d^3x ~ \tc^\dg_\s \vA^T \cdot \nabla \tc_\s \non \\
         \Hd &=& {e^2 \over 2m} \int d^3x ~ \tc^\dg_\s (\vA^T \cdot \vA^T) \tc_\s        \ .
\eea
Let 
\beq
            \D = |\D| e^{i\eta} \ .
\eeq
Then the ground state of $H_0$ is the usual BCS ground state
\bea
          \Psi^0_{BCS} &=&  \N_{BCS} \prod_{k} [u_k - v_k  e^{i\eta} \tc^\dg_\ua(k) \tc^\dg_\da(-k)] |0\rangle \ ,
\label{BCSvac}
\eea
where
\bea
           u(k) &=& \sqrt{\oh\left(1+ {\e_k\over E_k}\right)}  ~~,~~
           v(k) = \sqrt{\oh\left(1- {\e_k\over E_k}\right)}  \non \\
           E_k &=& \sqrt{\e_k^2 + |\D_k|^2}  ~~~,~~~ \e_k = {k^2\over 2m} - \e_F  \ ,
\eea
and self-consistency gives the gap equation
\beq
           \D  =  {g \over V} {\sum_k}' {\D \over 2\sqrt{\e_k^2 + |\D|^2}}  \ .
\label{gapeq}
\eeq
The self-consistency condition does not, however, fix the phase angle $\eta$, which is conventionally set to $\eta=0$.   

\subsection{Coulomb energy}

    The ions are treated here as static sources of charge $+pe$ with integer $p$, located at fixed positions
$\vx_n = 1,2,...,N_{ions}$.  We then make the following ansatz for the approximate ground state wave functional, which satisfies the physical state condition:
\beq
            \Psi_0  =  \Psi^0_{BCS}  \Psi_{ions} \Psi_A \ ,
\label{ground}
\eeq
where 
\beq
\Psi_{ions} = \exp\left[ip\sum_n \g_{\vx_n}\right]  \ .
\eeq
It is straightforward to verify  that $\Psi_0$ satisfies the physical state condition \rf{physical} with charge density \rf{chargedensity}, due to the inclusion of the $e^{i\g_x}$ phase factors in $\Psi^0_{BCS}$ and $\Psi_{ions}$.  We also find that inclusion of these factors leads to the energy due to Coulomb interactions among electrons and ions
\bea
\E_C &=& \langle \Psi_0 | \oh \int d^3x ~ E_L^2(x) | \Psi_0\rangle  \non \\
&=& \E_C^{electrons} + \E_C^{ions} + \E_C^{mixed} \ ,
\label{coulomb_energy}
\eea
where
\bea
\E_C^{electrons} &=& \oh \left( {e \over V} \sum_q {2 v^2_q \over u_q^2 + v_q^2} \right)^2 \int d^3x \int d^3y
           {1 \over 4\pi |\vx - \vy|} \ .  \non \\
\eea
The factor in parenthesis is essentially a position-independent charge density, so that $\E_C^{electrons}$ amounts to the Coulomb energy of a uniformly charged fluid in volume $V$.  We also find
\bea
\E_C^{ions} &=& \oh p^2 e^2 \sum_{n_1 \ne n_2}  {1 \over 4\pi |\vx_{n_1} - \vx_{n_2}|}  \non \\
\E_C^{mixed} &=& -pe \left( {e \over V} \sum_q {2 v^2_q \over u_q^2 + v_q^2} \right) \sum_n \int d^3x
           {1 \over 4\pi |\vx - \vx_n|} \ .
\eea
In these expressions we have dropped the singular self-interaction contributions, which can only be
treated correctly in the framework of a fully relativistic quantum field theory.

\subsection{Symmetry breaking and gauge fixing}

   Note that $\langle \C \rangle \propto e^{i\eta}$ in the BCS ground state \rf{BCSvac}, but $\eta$ is arbitrary.
This is of course the standard situation in spontaneous symmetry breaking.  Formally, one needs to add
an explicit symmetry breaking term such as
\beq
          - J \sum_k [e^{-i\eta_0} \tc_\da(-k) \tc_\ua(k) + e^{i\eta_0} \tc^\dg_\ua(k) \tc_\da^\dg(-k)]
\eeq
to the Hamiltonian \rf{HBCS0}, and take the infinite volume limit followed by the $J\ra 0$ limit.  This results in 
$\langle \C \rangle \propto e^{i\eta_0}$ with some definite phase angle $\eta_0$.
In fact this procedure is really shorthand for the fact that in the absence of any explicit symmetry breaking, in a finite
volume and finite temperature, $\langle \C \rangle$ will be non-zero at any given time, with a phase that fluctuates slowly in time, averaging to zero over a long time period.  The rate of variation in the phase, however, goes to zero as the volume tends to infinity.

   We now observe that under the global transformation $\th(x)=\th$, the Cooper pair operator transforms as 
 \beq \C \ra e^{2i\th} \C \eeq
by virtue of \rf{global}.  Then, since $\langle \C \rangle \ne 0$, it would seem that a global subgroup of the gauge symmetry is broken,\footnote{The symmetry which is actually broken is the quotient group $U(1)/Z_2$, since $\C$ is invariant under the gauge transformations $e^{i\th}=\pm 1$} and the first question is how this gauge symmetry breaking can be consistent with the Elitzur Theorem.  The answer is that Elitzur's Theorem does not apply to global transformations of any kind.  Consider an operator $Q$ which is non-invariant under a gauge transformation carried out in a fixed finite volume $V_Q$.  Elitzur's theorem states that $\langle Q \rangle = 0$ even when we carry out the usual procedure of adding a term to break the symmetry, take the infinite volume limit, and then remove the breaking.   What is crucial, however, is that $Q$ will vary under a local gauge transformation carried out only in a fixed volume, which remains fixed
in the thermodynamic limit.  What Elitzur showed is that in this situation the effect of the breaking term can be bounded, even in the infinite volume limit, by some small parameter which is taken to zero at the end.  Details can be found in \cite{Elitzur:1975im,Itzykson:1989sx}.  If, on the other hand, $Q$ only varies under transformations carried out at every
site on the lattice (or, in the present case, throughout the volume of the solid), the bound fails in the thermodynamic limit, and the theorem does not apply.  And of course it must fail in this situation, for otherwise Elitzur's argument would also rule out the spontaneous breaking of ordinary global symmetries. 
               
    But what sort of operator can vary only under a global subgroup of the gauge transformation, and not under any other
transformation in the gauge group?  The answer is that in general such operators can be associated with gauge fixing, and this
in turn means that the operators are completely non-local.  Let $Q_x$ be a local operator, and let $G[x;A]$ be the gauge
transformation taking $Q_x$ into some particular gauge, e.g.\ Coulomb gauge, which leaves unfixed a global subgroup of the gauge symmetry, i.e.\ the group of transformations $G(x)=G$.  Then
\beq
            \widetilde{Q}_x = G[x;A] \circ Q_x
\eeq
is invariant under all elements of the gauge group {\it except} the subgroup of global tranformations.  But now $\widetilde{Q}$ 
is a non-local operator.  In one particular gauge (e.g.\ Coulomb gauge) we will have $\widetilde{Q}=Q_x$.  This looks local, but it is not, because the gauge fixing itself is a non-local operation, acting at every point in the spatial volume.   

    We now recognize that the Cooper pair operator $\C$ which detects a breaking of the global symmetry is an operator of
exactly this type, because the U(1) transformation used to define the $\tc, \tc^\dg$ operators, i.e.
\beq
G[x;A] = e^{i\g_x} \ ,
\eeq
is precisely the transformation which brings the gauge and matter degrees of freedom into Coulomb gauge, as we
see from
\bea
           A_k(x) &\ra& A_k(x) - {1 \over e} \pa_k \g_x  \non \\
                       &\ra& A^T_k(x) \ .
\eea
Therefore, in Coulomb gauge only,
\bea
          \C &=& {\sum_k}'  \tc^\dg_\ua(k) \tc^\dg_\da(-k) \non \\
               &=& {\sum_k}'  c^\dg_\ua(k) c^\dg_\da(-k) \ .
\eea
Evaluating the expectation value of the non-local operator $\C$ is the same as dropping the $e^{i\g_x}$ factors, and
evaluating the resulting operator in Coulomb gauge.

     But Coulomb gauge is not unique in leaving unfixed a remnant global gauge symmetry.  Axial gauge, temporal gauge,
light-cone gauge and covariant gauges also have this property.  We could just as well evaluate $\C$ in any one
of those gauges, and perhaps find a non-zero expectation value.  Or perhaps not. The point we will make in a later section is
that this non-locality in the order parameter, equivalent to a choice of gauge, introduces a certain ambiguity into the concept of ``spontaneous breaking'' of a gauge symmetry, even when that symmetry is global, rather than local.

\subsection{Goldstone modes and the superconductor phase}

   Let us recall a simple derivation of the Goldstone theorem, which can be found in standard textbooks 
\cite{Zee:2003mt,*Schwartz:2013pla}.  Suppose we have an operator $Q$ (a ``charge'' operator) 
which commutes with the Hamiltonian, and that the set of transformations $\exp[i\th Q]$ is a U(1) symmetry group (possibly a subgroup of a larger symmetry), and $Q\Psi_0 \ne 0$.   We also suppose that $Q$ is associated with a conserved current and can be expressed as the spatial integral of a charge density
\beq
         Q = \int d^3x ~ J_0(x) \ .
\eeq
Because $[H,Q]=0$, the state $Q \Psi_0$ has the same energy as $\Psi_0$, namely the ground state energy $E_0$.
Now consider a state with momentum $\vk$
\beq
          |\vk \rangle = \int d^3x ~ e^{-i\vk \cdot \vx} J_0(x) |\Psi_0 \rangle \ .
\eeq
As $\vk \ra 0$, the energy of this state converges to the energy of $Q \Psi_0$, which is $E_0$.  The conclusion is that the excitation energy $E_{ex}(k)$ of state $|\vk \rangle$ above the ground state energy vanishes as 
$\vk \ra 0$, i.e.\ there exist gapless (or, in particle physics language, massless) excitations.  This is the Goldstone theorem.  

    Perhaps surprisingly, this argument correctly predicts the existence of gapless excitations in the normal state,
which is generally not considered to be a state of broken symmetry.  It may be of interest to see explicitly how this works
out in the normal state, and how that conclusion is evaded in superconducting state.
In the present case it is the number operator
\beq
           N = \int d^3x ~ c^\dg_\s(x) c_\s(x) = \int d^3x ~ \tc^\dg_\s(x) \tc_\s(x)
\eeq
which commutes with the Hamiltonian, and in fact $eN$ is the electric charge operator.  It is easy to see
that 
\beq
           [e^{i\th N},\tc^\dg_\s(x)] =  e^{i\th} \tc^\dg_\s(x) \ ,
\eeq
and as a consequence, operating on the BCS ground state
\bea
            e^{i\th N} \Psi^0_{BCS} &=&    e^{i\th N} \prod_k \Bigl(u_k + v_k \tc^\dg_\ua(\vk) \tc^\dg_\da(-\vk) \Bigr) |0 \rangle \non \\
                                                 &=&  \prod_k \Bigl(u_k + v_k e^{2i\th} \tc^\dg_\ua(\vk) \tc_\da^\dg(-\vk) \Bigr) |0 \rangle \ ,
\eea
which we recognize as a global U(1) gauge transformation $\th(x)=\th$ acting on the ground state.


    Let us ignore for the moment the transverse photon and ion degrees of freedom, and go back
to the usual BCS Hamiltonian
\bea
           H &=& \int d^3 x ~ c^\dg_\s(x) \left[ {1\over 2m}(-\nabla^2) - \e_F \right] c_\s(x) \non \\
           &-& {g\over V} {\sum}'_k {\sum}'_{k'} c^\dg_\ua(k) c^\dg_\da(-k) c_\da(k') c_\ua(-k') - \E_{grd} \ , 
 \label{Hgold}
\eea
where $\E_{grd}$ is the ground state energy, so that ${H|\Psi_0 \rangle = 0}$.  
$H$ is still invariant under the global transformations 
${c_\s(x) \ra e^{i\th} c_\s(x) , ~ c^\dg_\s(x) \ra e^{-i\th}c^\dg_\s(x)}$, and commutes with the generator of
those transformations, i.e.\ the number operator $N$
\bea
         N &=& \int d^3x ~ J_0(x)  \non \\
         J_0(x) &=& c^\dg_\s(x) c_\s(x)     \ .
\label{numop}
\eea                                              
Let us define
\bea
            N_q &=&   \int d^3x ~ c^\dg_\s(x) c_\s(x) e^{-i \vq \cdot \vx}  \non \\
                       &=& \sum_k c^\dg_\s(\vk) c_\s(\vk +\vq)  \ ,
\label{Nq}
\eea
so in this case
\beq
             |q \rangle = N_q |\Psi_0 \rangle  \ .
\label{state_q}
\eeq
Note that since $N_q$ cannot change electron number, any excitations above the ground state must correspond
to the creation of a particle-hole pair.  Introducing the usual Bogoliubov quasiparticle operators
\bea
         c_\ua(\vk) &=& u_k a_\ua(\vk) - v_k a^\dg_\da(-\vk) \non \\
         c_\da(-\vk) &=& v_k a^\dg_\ua(\vk) + u_k a_\da(-\vk) \non \\
         c^\dg_\ua(\vk) &=& u_k a^\dg_\ua(\vk) - v_k a_\da(-\vk) \non \\
         c^\dg_\da(-\vk) &=& v_k a_\ua(\vk) + u_k a^\dg_\da(-\vk)  \ ,
\label{quasi}
\eea
with the property that $a_\s(k) \Psi_0 = 0$, we have for $\vq \ne 0$
\bea
|q \rangle &=& \sum_k u_k v_{k+q} \{ a^\dg_\da(\vk) a^\dg_\ua(-\vk-\vq)  \non \\
                & & \qquad - a^\dg_\ua(\vk) a^\dg_\da(-\vk-\vq) \}  |\Psi_0 \rangle \ ,
\eea
and from here on our notational convention for momentum subscripts is that $u_{k \pm q}, v_{k \pm q},E_{k \pm q}$ means $u_{|\vk \pm \vq|},v_{|\vk \pm \vq|},E_{|\vk \pm \vq|}$
respectively.  We find the norm
\beq
\langle q | q \rangle = 2 \sum_k ( u_k^2 v_{k+q}^2 + u_k u_{k+q} v_k v_{k+q} ) \ ,
\eeq
and then evaluate $\langle q |H| q \rangle$ in the mean field approximation, replacing $H$ in \rf{Hgold} by
\beq
      H_{mf} = \sum_k E_k a^\dg_\s(k) a_\s(k)  \ ,
\eeq
which leads, in this approximation, to
\bea
   \E_q &=& { \langle q |H_{mf}| q \rangle \over \langle q | q \rangle }  \non \\
            &=&  { \sum_k (E_k+E_{k+q})  (u_k^2 v_k^2 + u_k u_{k+q} v_k v_{k+q} ) 
                    \over \sum_k ( u_k^2 v_{k+q}^2 + u_k u_{k+q} v_k v_{k+q} ) }  \ .
\eea
  
     Now in the normal phase, $\D_k=0$, we have $u_k v_k = 0$ for all $\vk$, and
$u_k v_{k+q} \ne 0$ only for $\vk$ and $\vk+\vq$ on opposite sides of the Fermi surface.  Then as $q \ra 0$,
the sum over $\vk$ is non-zero only in the immediate region of the Fermi surface, where $E_k=0$.
This means that $\E_q \ra 0$ as $\vq \ra 0$, i.e.\ there are gapless excitations in this phase which follow from
application of the Goldstone argument, whether or not one cares to describe this case as a phase of 
spontaneously broken symmetry.  

In the superconducting phase, $\D_k \ne 0$, the situation is different.  In this case $u_k v_k$ and $u_k v_{k+q}$
are non-zero, as $\vq \ra 0$, for $k$ roughly in the range $|\e_k| < \omega_D$, and in this range
$E_k, E_{k+q} > \D$.  Hence $\E_q \approx 2\D$ as $q \ra 0$, and there are no gapless excitations.  Excited
states (quasiparticle pairs) have a minimum energy of $2\D$.

    But this raises a question, since at $q=0$ {\it exactly} we must have $\E_q=0$.  This is because 
$|q=0 \rangle = N |\Psi_0 \rangle$, and since $[H,N]=0$ it must be that $|q=0 \rangle$ has the same
energy as the ground state, i.e.\ $\E_0 = 0$, not $\E_0 \approx 2\D$.  The apparent paradox is resolved by the
realization that exactly at $q=0$ there is an additional contribution to $N_q$ which does not annihilate the
ground state, namely
\bea
     \sum_k  v_k^2 a_\s(k) a_\s^\dg(k) = \sum_k v_k^2 (2 + a^\dg_\s(k) a_\s(k))  \ .
\eea
Redoing the calculation including these contributions, we have for the norm
\bea
  \langle q=0| q=0 \rangle &=& 4\left(\sum_k v_k^2\right)^2   + 4 \sum_k u_k^2 v_k^2 \ .
\eea
The first term on the right hand side is proportional to the square of the number of electrons in the system, i.e.\ to the square of the volume, while the second term grows only linearly with volume, and in addition 
only momenta in the neighborhood of the Fermi surface contribute to the second sum.  Therefore, up to O($1/V$)
corrections,
\bea
  \langle q=0| q=0 \rangle &=& 4\left(\sum_k v_k^2\right)^2 \ .
\eea
Then, since $H_{mf} |\Psi_0 \rangle = 0$, we have  
\bea
   \E_0 &=& { \langle q=0 |H_{mf}| q=0 \rangle \over \langle q=0 | q=0 \rangle }  \non \\
            &=&  { \sum_k 2 E_k u_k^2 v_k^2  
                    \over \left(\sum_k v_k^2\right)^2  } \non \\
             &=&  0 + \text{O}(1/V)  \ ,
\eea
where the last line follows since the numerator in the second line is O($V$), while the denominator is O($V^2$).
The fact that $\E_0$ is not exactly zero, but differs from zero by a term of order $1/V$, can be attributed to the mean field approximation, which in the BCS case is also only accurate up to corrections of this order.

   So we have seen that the textbook argument \cite{Zee:2003mt,*Schwartz:2013pla} can be applied to both the normal and superconducting phases.  The normal phase has gapless excitations in accordance with this argument. The superconducting phase, however, evades this conclusion in an interesting way, via a discontinuity in $\E_q$
(the energy of the low momentum $|q \rangle$ state) precisely at $q=0$.\footnote{In an insulator the proof is
evaded by the fact that there are no small $q$ particle-hole excitations near the Fermi surface, as $N_q$ in
\rf{Nq} annihilates the ground state for small $q$.  Hence there is no smooth limit to $q=0$.}

The  superconducting and normal phases are of course distinguished by the
expectation value of the Cooper pair operator $\tc_\ua^\dg(k) \tc_\da^\dg(-k)$, which vanishes in the normal phase
and is non-zero in the superconducting phase.\footnote{Actually this operator also vanishes in a system with a definite
number of electrons.  In that case one may consider instead correlators such as
$\langle c_\ua(x) c_\da(x) c_\ua^\dg(y) c_\da^\dg(y) \rangle$ in the limit of large separation $|x-y|$, which would still
vanish in the normal phase and be non-zero in the superconducting phase.}  However, the simple model described
by the Hamiltonian in \rf{Hgold} has no coupling to gauge fields, and no local gauge invariance.  In a gauge theory,
order parameters such as the Cooper pair creation operator in the BCS theory, or the charged scalar field in
the Ginzburg-Landau effective theory, transform under local gauge transformations.  Hence their expectation
values vanish unless either (i) the gauge is fixed; or (ii) we employ a construction which is equivalent to gauge fixing, as explained in part A of this section.  As we will see in section \ref{sec4}, this introduces an ambiguity, in the
sense that the vanishing or finiteness of the order parameter turns out to be gauge dependent.

\section{\label{sec3} Origin of the transverse photon mass}

Let $\Psi_{BCS}$ represent the ground state up to first order corrections in $e$
\bea
       \Psi_{BCS} &=&  \Psi^0_{BCS} +  \sum_{n\ne 0} { \langle \Phi_n|\Hp|\Psi^0_{BCS} \rangle  \over E_0 - E_n} \Phi_n \non \\
                          &=&  \Psi^0_{BCS} +  \Psi^1_{BCS}  \ ,
\label{PBCS1}
\eea    
where the $\Phi_n,E_n$ are excited energy eigenstates and eigenvalues of $H_0$, with $E_0$ the ground state energy.  The derivation of the London equation from this state goes back to the original BCS paper \cite{PhysRev.108.1175}. But to see the origin of photon mass, or indeed to have photons of any kind,  it is necessary to work with a quantized electromagnetic field, and Hamiltonian $H_A$.   The fermionic operator algebra is, however, essentially identical to the algebra used to derive the London equation.

   Altogether, including the first order corrections in \rf{PBCS1}, we have a ground state of the form
\beq 
          \Psi_{all} = \Psi_{BCS} \Psi_{ions} \Psi_A  \ .
\eeq
In $\Psi_A$ we must allow for the possibility of a photon mass.  The ground state of the free Maxwell field is
\beq
            \Psi_A^{free} \sim  \exp\left[ - \oh \sum_k |k| A_i^T(k)  A_i^T(-k) \right]  \ .
\eeq
On the other hand, the ground state of a free massive scalar field $\phi$ of mass $m$ is
\beq
          \Phi \sim \exp[\left[ - \oh \sum_k \sqrt{k^2 + m^2} \phi(k) \phi(-k)  \right] \ ,
\eeq
which motivates the ansatz
\bea
       \Psi_A(\k) &=&  \N_A \exp\left[ - \sum_k \sqrt{k^2 + \m^2(k)}  A_i^T(k)  A_i^T(-k) \right] \non \\
       &=& \N_A \exp\left[ - \sum_k \sum_\l \sqrt{k^2 + \m^2(k)}  A(k,\l)  A(-k,\l) \right]  \  . \non \\
\eea
where $\m(k)$ is a momentum-dependent photon ``mass'' of some kind, with $\N_A$ the normalization constant.  
The expression $\m(k)$ is a variational term, chosen to minimize the vacuum energy
\beq
           \E =  {\langle \Psi_{all} | H_{BCS}^{mf} + H_A  | \Psi_{all} \rangle \over \langle \Psi_{all} | \Psi_{all} \rangle} \ .
\eeq
Define
\bea
        \E_0 &=& \langle \Psi^0_{BCS} \Psi_A \Psi_{ions} | H_0 + H_A  | \Psi^0_{BCS} \Psi_A \Psi_{ions} \rangle \non \\
                &=&  E_0 + \E_A + \E_C \non \\
        \E_n &=& \langle \Phi_n \Psi_A \Psi_{ions} | H_0 + H_A | \Phi_n \Psi_A \Psi_{ions} \rangle \non \\
                &=&  E_n + \E^0_A +  \E_{Cnn}  \ ,
\eea
where $\E_C$ is the O($e^2$) Coulomb energy in \rf{coulomb_energy}, $E_0=0$ since the ground state energy
is subtracted in the definition of $H_0$, and 
\bea
             E_n &=& \langle \Phi_n | H_0 | \Phi_n \rangle \non \\
           \E_A &=& \langle \Psi_A | H_A | \Psi_A \rangle \non \\  
           \E'_{Cnn} &=& \left\langle \Phi_n \Psi_{ions} \left| \oh \int d^3x ~{\bf E}_L^2 \right| \Phi_m \Psi_{ions} \right\rangle  \ .
\eea
$\E'_{Cnn}$ is an additional Coulomb energy associated with creation of a few quasiparticles above the ground state.  Since this will contribute to the total energy only at O($e^4$), we can neglect it.   Also define
\bea
           \Psi^0_{all}  &=&  \Psi^0_{BCS}  \Psi_{ions} \Psi_A \non \\
           \Psi^1_{all}  &=&  \Psi^1_{BCS}  \Psi_{ions} \Psi_A  \ .
\eea
Then
\begin{widetext}
\bea
\E &=& {1\over  1 +  \langle \Psi^1_{all}| \Psi^1_{all} \rangle } \left\{\E_0 +  \langle \Psi^1_{all}| H_{BCS}^{mf} + H_A |\Psi^1_{all} \rangle +  \langle \Psi^0_{all}| \Hd |\Psi^0_{all} \rangle  \right.  \non \\
     & & \qquad \qquad \qquad \left.  + \langle \Psi^1_{all}| \Hp |\Psi^0_{all} \rangle +  \langle \Psi^0_{all}| \Hp |\Psi^1_{all}\rangle \right\} \ .
\eea
Keeping only terms up to O($e^2$),
\bea
     \E &=& \E_0(1 - \langle \Psi^1_{all}| \Psi^1_{all} \rangle) +  \langle \Psi^1_{all}| H_0+H_A|\Psi^1_{all} \rangle +
      \langle  \Psi^0_{all}| \Hd |\Psi^0_{all} \rangle + \langle \Psi^1_{all}| \Hp |\Psi^0_{all} \rangle 
      +  \langle \Psi^0_{all}| \Hp |\Psi^1_{all} \rangle \non \\
      &=& \E_0 +  \left\langle \Psi_A \left|  \sum_{n\ne 0} (- \E_0 + \E_n)  { |\langle \Phi_n|\Hp|\Psi^0_{BCS} \rangle|^2  
              \over (E_0 - E_n)^2}     +   2\sum_{n\ne 0}{ |\langle \Phi_n|\Hp|\Psi^0_{BCS} \rangle|^2  
              \over (E_0 - E_n)}  + \langle \Psi^0_{BCS} | \Hd |\Psi^0_{BCS} \rangle  \right| \Psi_A \right\rangle \ .
\eea
\end{widetext}
Up to O($e^2$) in $\E$ we may set $\E_n - \E_0 = E_n-E_0$, leaving
\bea
  \E &=& \E_0 +  \bigg\langle \Psi_A \bigg| \langle \Psi^0_{BCS} | \Hd |\Psi^0_{BCS}\rangle  \non \\
       & &  -  \sum_{n\ne 0}{ |\langle \Phi_n|\Hp|\Psi^0_{BCS} \rangle|^2  \over E_n - E_0}    \bigg| \Psi_A \bigg\rangle \ .
\label{Eslash}
\eea             
It is convenient to evaluate the right hand side of \rf{Eslash} with the help of the Bogoliubov quasiparticle operators \rf{quasi}
which diagonalize $H_0$, and then a little operator algebra gives
\bea
           T_1 &\equiv& \sum_{n\ne 0}{ |\langle \Phi_n|\Hp|\Psi^0_{BCS} \rangle|^2  \over E_n - E_0}    \non \\
                  &=& {1\over V} {e^2 \over m^2} \sum_p \sum_q {(u_p v_{p-q} - u_{p-q} v_p)^2 \over E_p + E_{p-q} } |\vp \cdot
                  A^T(q) |^2 \non \\
                 &=& \sum_q r(q) |A^T(q) |^2 \ ,
\label{T1}
\eea
and
\bea
            T_2 &\equiv&  \langle \Psi^0_{BCS} | \Hd |\Psi^0_{BCS}\rangle  \non \\
                    &=& {e^2 \over 2m} {1\over V} \sum_p 2 v_p^2 \sum_q |A^T(q)|^2  \non \\
                    &=& {e^2 \over 2m} \eta \sum_q  |A^T(q)|^2  \ ,
\label{T2}
\eea 
where $\eta = N/V$ is the electron number density.  The part of the energy $\E_\m$ which is due to quantum fluctuations of the transverse $A$-field, and which depends on the photon mass term $\m(q)$ is then
\bea
          \E_\m &=&  \bigg\langle \Psi_A \bigg| H_A + \sum_q \left({e^2 \over 2m} - r(q^2) \right) 
             A^T(q) |^2 \bigg| \Psi_A \bigg\rangle \non \\ 
                    &=& \oh \sum_q \left( \sqrt{q^2 + \m^2(q)} + {-\m^2 + {e^2 \over 2m} \eta 
                    -r(q) \over 2\sqrt{q^2 + \m^2(q)} } \right) \ .
\eea
This energy is minimized by the momentum-dependent photon mass
\beq
          \m(q) = \sqrt{ {e^2 \over 2m} \eta -r(q) } \ .
\label{muq}
\eeq
Although $\D$ has been fixed here from the start by the self-consistency equation \rf{gapeq}, it could equally well be treated as a variational parameter, and determined by minimization of the ground state energy, cf.\ \cite{Timm}.  This approach yields the same equation \rf{gapeq} for $\D$.

    In the original BCS paper it was shown that the expectation value of the current in the state $\Psi_{BCS}$ had a 
paramagnetic and diamagnetic component, which closely correspond to the $r(q)$ and ${e^2 \over 2m} \eta$ terms
respectively, and that for the paramagnetic term $r(q) \ra 0$ in the $q \ra 0$ limit.  This is obvious by inspection of the
second line in eq.\ \rf{T1}.  However, they also showed that the paramagnetic term cancels the diamagnetic term in
a normal metal, in the same limit, which means that $\m(0)=0$.  This is a little less obvious, so for completeness we
re-derive this cancellation at $\D=0$ in the Appendix.

 For the superconducting state, with $\D \ne 0$, we have ${r(0)=0}$, and the photon mass is $\m(q=0) = 
 \sqrt{e^2 \eta/2m}$.  At non-zero momenta in the superconducting the state, the general expression is
\bea
   r(q) &=&  {1\over (2\pi)^3} {e^2 \over m^2} \int d^3p {(u_p v_{p-q} - u_{p-q} v_p)^2 p_x^2 \over
                                 E_p + E_{p-q} } \ ,
\eea
which must be evaluated numerically .
    
\vspace{30pt}

\section{\label{sec4} The Ambiguity of Spontaneous Gauge Symmetry Breaking} 

   As explained in section \ref{sec2}, an observable which transforms only under a global subgroup of a local gauge
symmetry may have a non-zero vacuum expectation value; this is not forbidden by the Elitzur theorem.  But is this
what is meant by a ``spontaneously broken'' gauge symmetry?  We believe this phrase is ambiguous,
for the simple reason that different operators, each of which transform under a global subgroup of the 
gauge symmetry, may not agree on exactly where in the phase diagram the symmetry is actually broken.  They may not even agree on whether the symmetry is broken at all.  It is now time
to elaborate on this point.  For this purpose we will focus on the abelian Higgs model, where the scalar field has
charge $q e$, where $q$ is an integer.  The abelian Higgs model at $q=2$ is a relativistic version of the Ginzburg-Landau
effective action with a lattice regularization and compact U(1) gauge group.  The quantum mechanical model is described by
\bea
          Z &=& \int DU_\m D\phi ~ e^{-S} \ ,
\eea
with action
\bea
          S &=& -\b \sum_{x} \sum_{\m < \n} \text{Re}[U_\m(x) U_\n(x+\hat{\n}) U^*_\m(x+\hat{\n}) U^*_\n(x)] \non \\
              & &      - \g \sum_{x} \sum_{\m=0}^3 \text{Re}[\phi^*(x) (U_\m(x))^q \phi(x+\hat{\m})] \ .
\label{ahiggs}
\eea
The gauge field is an element of the U(1) group, i.e.\ $U_\m(x) = e^{i\chi_\m(x)}$, and for simplicity we also take
the Higgs field to have unit modulus, i.e.\ $\phi(x) = e^{i\d(x)}$.  Finite temperature is imposed by a finite extension $N_t$ of
the lattice in the time direction, i.e.\ $T = 1/(N_t a)$, where $a$ is the lattice spacing.  The action is
invariant under U(1) gauge transformations
\bea
          U_\m(x) \ra U'_\m(x) &=& e^{i\th(x)} U_\m(x) e^{-i\th(x+\hat{\m})} \non \\
          \phi(x) \ra \phi'(x) &=& e^{i q \th(x)} \phi(x)  \ .
\label{gtrans_ah}
\eea

\begin{figure}[h!]   
\centerline{\includegraphics[scale=0.7]{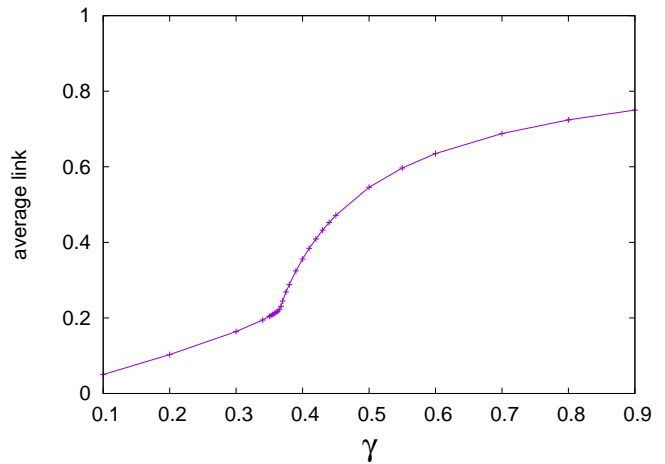}}
\caption{Average link $L$ (eq.\ \rf{L})  vs.\ $\g$ at $\b=2.0$ on a $16^4$ lattice volume.  The transition from the
massless to the Higgs phase is located at $\g=0.365$, where the slope changes abruptly.}
\label{kink}
\end{figure}

\begin{figure}[h!]   
\centerline{\includegraphics[scale=0.7]{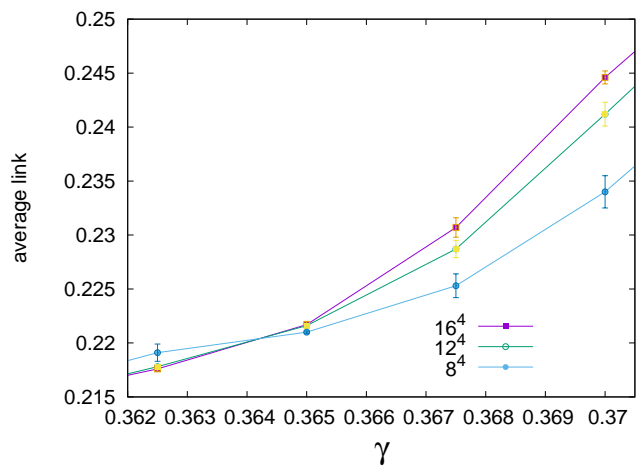}}
\caption{Closeup of the $L$ vs.\ $\g$ data at $\b=2.0$ in the immediate neighborhood of the transition, on $8^4, 12^4, 16^4$ lattice volumes.  Note that the change in slope at the transition near $\g=0.365$ becomes more abrupt with
increasing volume.}
\label{kink1}
\end{figure}

\begin{figure}[h!]   
\centerline{\includegraphics[scale=0.7]{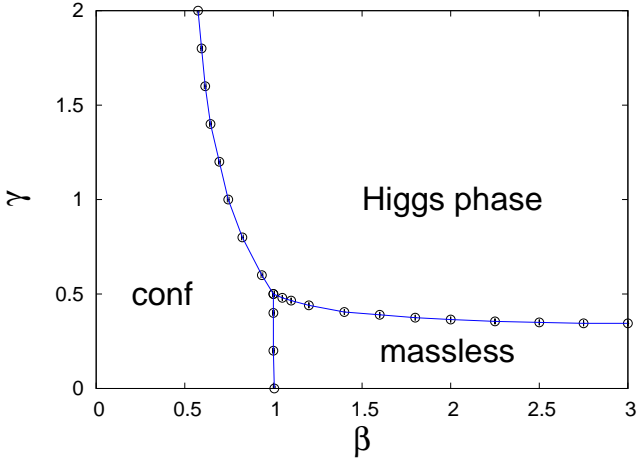}}
\caption{The $q=2$ phase diagram.  The confining, massless, and Higgs phases are completed separated
by thermodynamic transitions.}
\label{thermo}
\end{figure}

    The phase diagram of the abelian Higgs model in the space of couplings $\b,\g$ and charges $q=1,2,6$ was
determined long ago by Ranft et al.\ \cite{Ranft:1982hf}, albeit on lattices which were tiny ($4^4$) by todays standards, with transition points located by a method (hysteresis curves) which has since been superseded 
by other methods.   For this article we have determined the transition points in the $q=2$ theory, from the confinement to the Higgs or massless phases, from the location of peaks in the plot of plaquette susceptibility vs.\  $\b$, at fixed $\g$, on a $12^4$ lattice volume.  Transition points from the massless to Higgs phase are located from the position of a ``kink,'' i.e.\ an abrupt change in slope, in a plot of the link action
\beq
           L = {1\over V} \sum_x \sum_\m \langle \text{Re}[\phi^*(x) U^2_\m(x) \phi(x+\hat{\m})] \rangle
\label{L}
\eeq
vs.\ $\g$.  An example of data for $L$ vs.\ $\g$ at $\b=2$, on a $16^4$ lattice, is plotted in Fig.\ \ref{kink}, and the kink
is apparent at ${\g \approx 0.365}$.  Since this behavior should reflect a non-analyticity of the free energy in the thermodynamic limit, we would expect the change in slope at the transition to become increasingly abrupt as the volume increases.  In Fig.\ \ref{kink1} we show our data for $L$ vs.\ $\g$ in the immediate neighborhood of the transition point,  at lattice volumes $8^4, 12^4, 16^4$, which agrees with this expectation.  

   In the end our results for the thermodynamic 
phase structure of the $q=2$ theory, displayed in Fig.\ \ref{thermo}, are not far off the
old results of \cite{Ranft:1982hf}.   We should point out that
in the confinement phase denoted ``conf'' in Fig.\ \ref{thermo}, what is really confined are test charges with $\pm 1$ units  ($q=1$) of electric charge.  The meaning of 
confinement in this region for $q=2$ charges, and how the confinement phase for $q=2$ charges is distinguished from the Higgs phase, is not at all trivial, and will be discussed in section \ref{CSc}.  The massless phase is continuously connected
to the massless phase of the pure gauge theory at $\g=0$, which is known to have a transition between the confined and massless phases at $\b=1$.
   
      Let us define, in the $q=2$ abelian Higgs theory, two different order parameters, $Q_L$ and $Q_T$, each of which transforms under a global subgroup, defined by $\th(x)=\th$, of the local U(1) gauge symmetry, via
\beq
           Q_L \ra e^{2i\th} Q_L ~~~,~~~  Q_T \ra e^{2i\th} Q_L \ ,
\eeq
but which are invariant under any local gauge transformation.  As explained previously, such operators can be defined
by a gauge choice which leaves unfixed a remnant global subgroup of the full gauge symmetry, i.e.
\beq
            Q_{L,T} = {1\over V} \sum_x G_{L,T}(x;U) \phi(x) \ ,
\label{QTL}
\eeq
where $G(x;U)$ is the gauge transformation which takes the gauge field into some gauge which leaves unfixed the remnant global symmetry, and $V$ is the lattice volume.  Let $L$ denote Landau gauge, which is the gauge that maximizes
\beq
           R_L = \sum_{x} \sum_{\m=1}^4 \text{Re}[U_\m(x)] \ ,
\eeq
and let $T$ denote ``maximal'' temporal gauge, in which 
\bea
          U_4(x) &=& 1 ~~\text{for} ~~ x_4 \ne 1 \non \\
          U_3(x) &=& 1 ~~\text{for} ~~ x_4 = 1, x_3 \ne 1  \non \\
          U_2(x) &=& 1 ~~\text{for} ~~ x_4 = 1, x_3 =1, x_2 \ne 1  \non \\
          U_1(x) &=& 1 ~~\text{for} ~~ x_4 = 1, x_3 =1, x_2 =1, x_1 \ne 1  \ .
\label{maxtemp}
\eea
Landau and maximal temporal gauge fix all but a remnant global symmetry $\th(x)=\th$.   

In lattice Landau gauge, however, we have to contend with the Gribov ambiguity, i.e.\ the fact that there are many local maxima of $R$, and therefore the full specification of
$G_{L}(x;U)$ depends on the Gribov copy selected.  Obviously no fully gauge invariant observable can depend
on such a choice, but we are dealing here with order parameters which, as we shall see, most definitely depend on the gauge.  The most natural choice in Landau gauge would be the transformation $G_{L}$ which brings $R$ to an absolute maximum.  Numerically this is impossible to achieve in practice, in fact the determination of the absolute maximum is believed to be NP hard.  However, any deterministic algorithm will select a unique gauge copy corresponding to a local maximum of $R$, given a particular lattice configuration $U_\m(x)$, so the specific gauge-fixing algorithm used by the computer may be regarded as part of the specification of the gauge choice.

   We may also define lattice Coulomb gauge as the gauge which maximizes
\beq
           R_C = \sum_{x} \sum_{i=1}^3 \text{Re}[U_i(x)] \ ,
\eeq
and $G_C(x)$ as the gauge transformation to Coulomb gauge.  In Coulomb gauge there remains a symmetry under
gauge transformations which depend only on time, i.e.\ $\th(\vx,t)=\th(t)$.  On any given time slice, this is a remnant
global symmetry, which may be spontaneously broken on that time slice.  We therefore define the $Q$ observable on
each time slice as 
\beq
            Q_{C}(t) = {1\over V_3} \sum_{\vx} G_{C}(\vx,t;U) \phi(\vx,t) \ ,
\label{QC}
\eeq
where $V_3$ is the $D=3$ dimensional spatial volume of the time slice.  Of course there is no true phase transition
on a finite volume, and so in practice we compute, in a fixed volume $V$
\bea
          Q_{L,T}(V) &=&  {1\over V} \left| \sum_x \phi(x) \right| \non \\
          Q_C(V_3,t) &=&  {1\over V_3} \left| \sum_\vx \phi(\vx,t) \right| \ ,
\eea
with $\phi(x)$ fixed to maximal temporal, Landau, or Coulomb gauge, respectively, and extrapolate the results to $V=\infty$.  Transitions are located by peaks in the susceptibilities
\bea
          \chi_L &=& V( \langle Q_L(V)^2 \rangle -  \langle Q_L(V) \rangle^2 ) \non \\
          \chi_C &=& {1\over N_t} \sum_{t=1}^{N_t} V_3 ( \langle Q_C(V_3,t)^2 \rangle -  \langle Q_C(V_3,t) \rangle^2 ) \ .
\eea

\begin{figure*}[t!]
\subfigure[~Landau gauge]  
{   
 \label{lan}
 \includegraphics[scale=0.65]{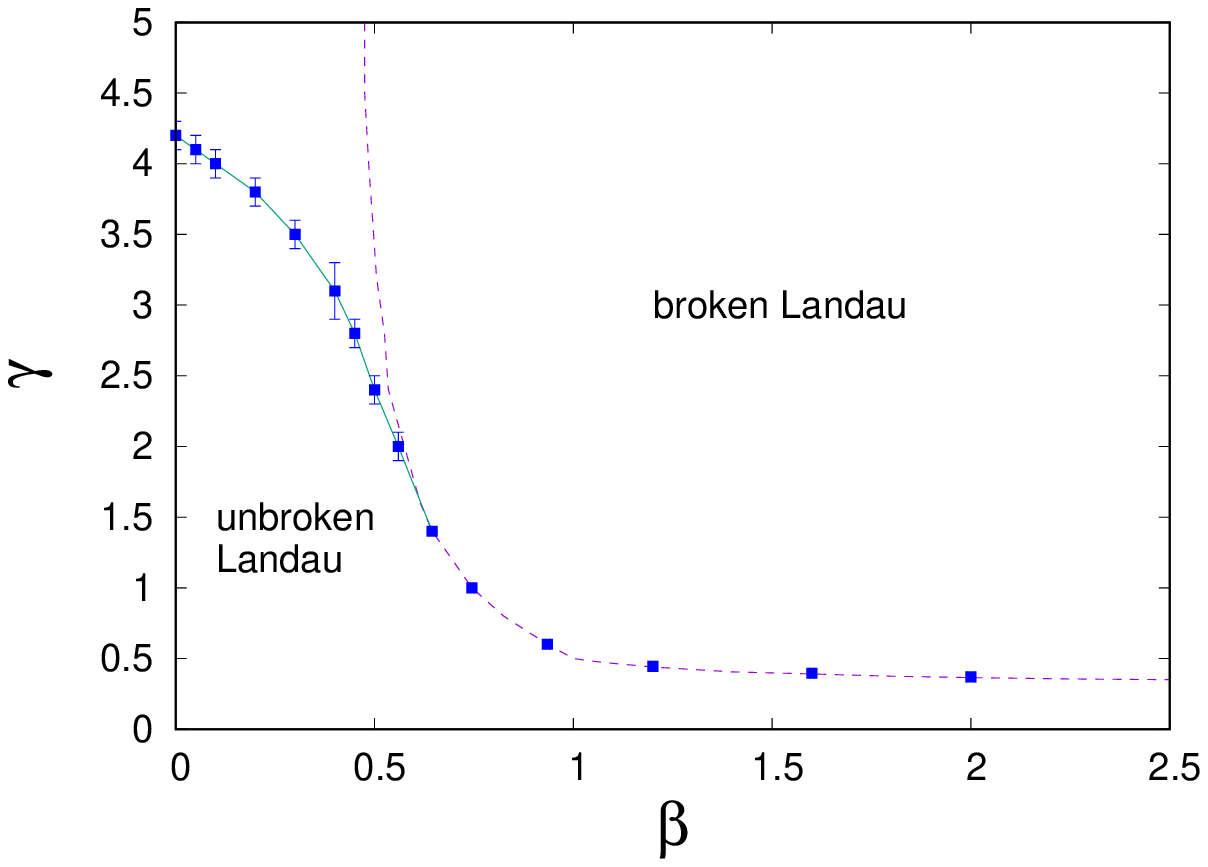}
}
\subfigure[~Coulomb gauge]  
{   
 \label{coul}
 \includegraphics[scale=0.65]{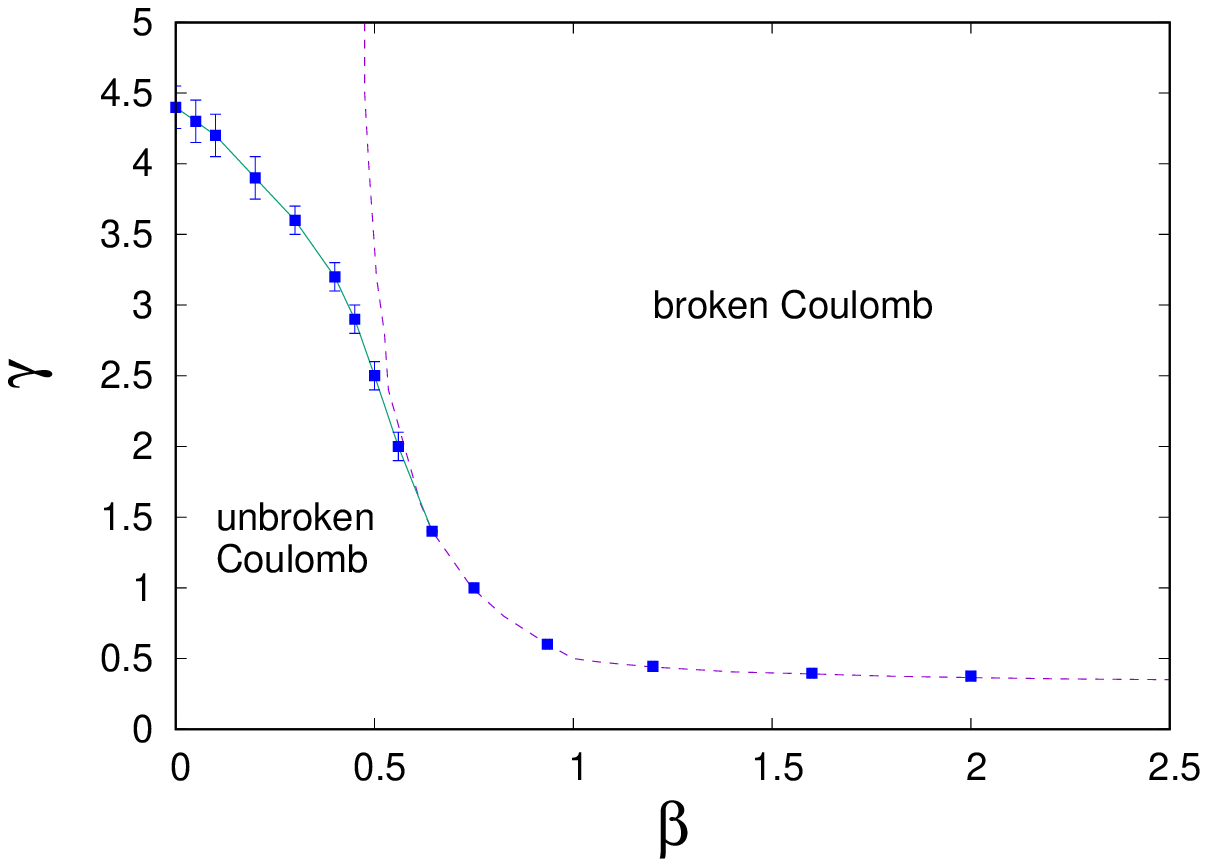}
}
\caption{Transition points for the breaking of a global remnant gauge symmetry in (a) Landau and (b) Coulomb gauges.
The dashed line is the line of thermodynamic transition shown in Fig.\ \ref{thermo}.} 
\label{remnant}
\end{figure*}

\begin{figure}[htb]
\includegraphics[scale=0.65]{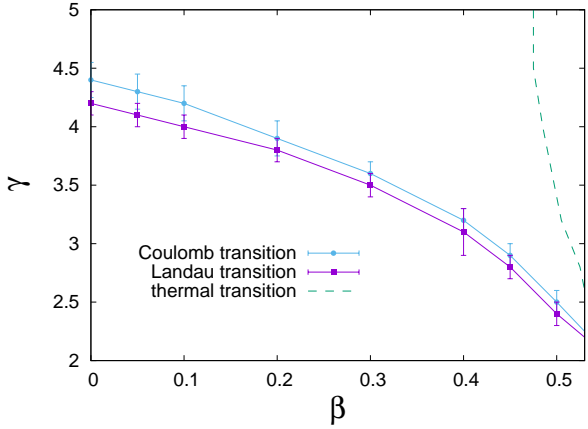}
\caption{Closeup of the remnant symmetry breaking points in Landau and Coulomb gauges, away from the
line of thermal transitions.}
\label{lancoul}
\end{figure}

   We have seen in Fig. \ref{thermo}  that in the $q=2$ case there are three phases, which we denote as ``massless,'' ``Higgs,'' and ``confinement,'' completely separated from one another by lines of thermodynamic transition.  In the massless phase all three of the order parameters $Q_L,Q_C,Q_T$  extrapolate to zero at infinite volume, as one might expect.  
Within the Higgs phase, the remnant global gauge symmetry is spontaneously broken in the full volume, for Landau
gauge, and in any time slice, in Coulomb gauge.
However, the remnant symmetries in Landau and Coulomb gauges are {\it also} broken inside the confinement phase,
at higher $\g$ values, and moreover the Landau and Coulomb transition lines do not coincide within the confinement
phase.  The phase diagrams for remnant symmetry breaking, for Landau and Coulomb gauges, are shown in 
Fig.\ \ref{remnant}.
In this figure the remnant symmetries break at the points shown, while the thermodynamic transition is indicated by
the dashed line.   We see that at small $\b$ there is a line of remnant symmetry breaking in the confined region which
does not correspond to any thermodynamic transition, and which lies entirely in the confined phase.   Moreover
the transition line in the confined phase is slightly different in Landau  and Coulomb gauges, as seen in Fig.\ \ref{lancoul}.  Already we can conclude that spontaneous breaking of remnant gauge symmetry is gauge dependent.  

\begin{figure}[h!]   
\centerline{\includegraphics[scale=0.6]{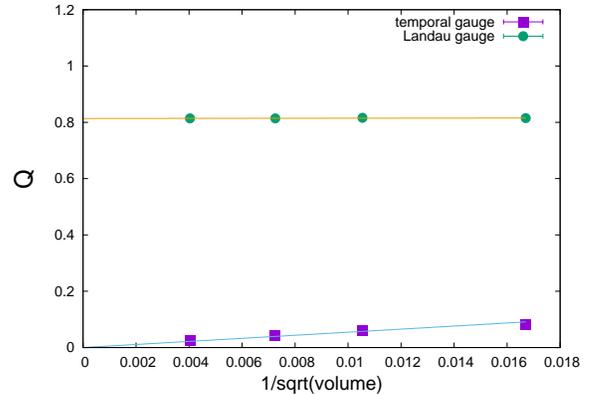}}
\caption{The Landau and temporal gauge order parameters $Q_L$ and $Q_T$ vs.\ inverse square root of the
lattice volume $1/\sqrt{V}$, inside the Higgs phase at $\b=1.2, \g=0.7$.  The remnant global gauge symmetry is
broken in this phase in Landau gauge, according to $Q_L$, but is not broken in temporal gauge, according to $Q_T$,
which extrapolates to zero at infinite volume.}
\label{temporal}
\end{figure}

    Even within the Higgs phase, spontaneous gauge symmetry breaking is not seen all gauges which leave unfixed a global subgroup of the gauge symmetry.  In Fig.\ \ref{temporal}
we display $Q_L$ and $Q_T$ vs.\ $1/\sqrt{V}$, at a point $\b=1.2, \g=0.7$ which is inside the Higgs phase (as determined by thermodynamic transitions, see Fig.\ \ref{thermo}).  It is seen $Q_T$ extrapolates to zero at infinite lattice
volume inside the Higgs phase, while $Q_L$ does not.  Here again we have evidence of the gauge dependence of spontaneous symmetry breaking of remnant gauge symmetry.

{
 
     The ambiguity outlined here is certainly not limited to the abelian Higgs model, in fact it was first noted in ref.\ 
\cite{Caudy:2007sf} for the SU(2) gauge-Higgs model, with the Higgs field in the fundamental representation of the gauge group.  The action in this case is
\bea
          S &=& - \b \sum_{x} \sum_{\m < \n} \oh \text{Tr}[U_\m(x) U_\n(x+\hat{\n}) U^\dg_\m(x+\hat{\n}) U^\dg_\n(x)] \non \\
              & &      - \g \sum_{x,\m} \oh \text{Tr}[\phi^\dg(x) U_\m(x) \phi(x+\hat{\m})] \non \\
              &=&  S_W + S_H \ ,
\label{SU2gh}
\eea
with $\phi(x)$ an SU(2)-valued field.  it was found that the breaking of the residual gauge invariance in Coulomb and Landau gauges occurs along different transition lines, shown in Fig.\ \ref{su2_break}.   There is no thermodynamic transition in the region of the phase diagram where the Landau and Coulomb lines differ.

\begin{figure}[h!]   
\centerline{\includegraphics[scale=0.65]{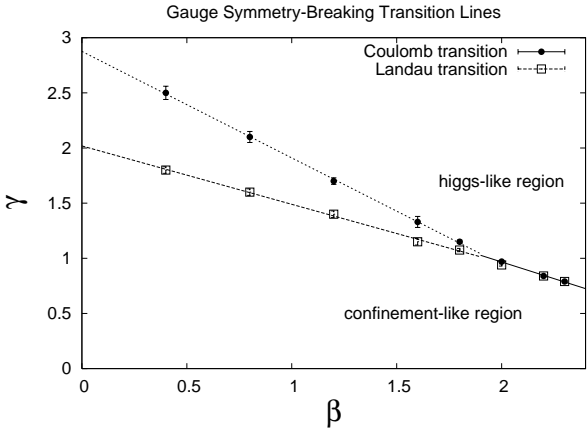}}
\caption{The remnant symmetry breaking lines for Coulomb and Landau gauges in SU(2) gauge-Higgs theory
with action \rf{SU2gh}.}
\label{su2_break}
\end{figure}      

\subsection{\label{z2sec} $Z_2$ symmetry breaking}

    Apart from gauge symmetry, the action of the $q=2$ gauge-Higgs model is invariant under the following symmetry
 \beq
           U_4({\bx},0) \ra z U_4({\bx},0)  ~~~ \mbox{all $\bx$ at $t=0$}
\label{z2}
\eeq
where $z = \pm 1$ is an element of the $Z_2$ group.  For pure gauge theory ($\g=0)$  $z$ is an element of U(1), and the symmetry is known as ``center symmetry.''  In the $q=2$ model the U(1) center symmetry is broken down to $Z_2$, while in the $q=1$ model the symmetry absent entirely.  A gauge invariant observable which transforms non-trivially
under the $Z_2$ symmetry is the Polyakov line
\beq
          P(\bx) = U_4(\bx,1) U_4(\bx,2) ... U_4(\bx,N_t)
\eeq
where  $P(\bx) \ra z P(\bx)$ under \rf{z2}.  Therefore the expectation value of the Polyakov line is an order parameter
for spontaneous breaking of global $Z_2$ symmetry.  Moreover, since 
\beq
     \langle P \rangle \sim e^{-F/kT} 
\eeq
where $F$ is the free energy of a static source with a single unit of charge, $\langle P \rangle = 0$ implies confinement,
and $\langle P \rangle \ne 0$ means non-confinement, of particles with a single unit ($q=1$) of charge.  Thus we
expect $\langle P \rangle = 0$ in the region labeled ``conf'' of the phase diagram shown in Fig.\ \ref{thermo}, and
$\langle P \rangle \ne 0$ in the Higgs and massless phases.  We have verified (on a $12^3 \times 6$ lattice volume) that the transition happens across the transition line shown in Fig.\ \ref{z2fig}, separating the confinement from the Higgs and massless phases.  

\begin{figure}[h!]   
\centerline{\includegraphics[scale=0.6]{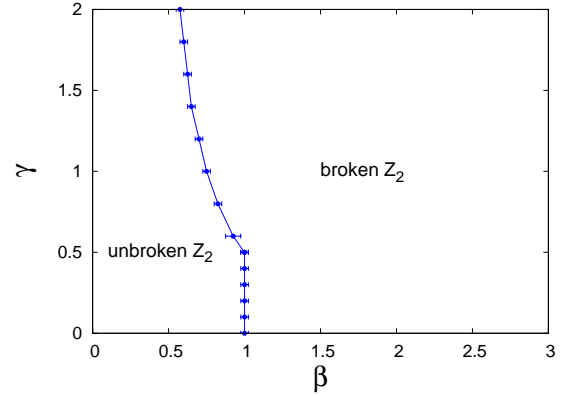}}
\caption{The $Z_2$ transition line, as detected by Polyakov lines on a $12^3 \times 6$ lattice.  This line
coincides with thermodynamic transitions from the confinement phase for $q=1$ (but not $q=2$) test charges, to the 
massless and Higgs phases.}
\label{z2fig}
\end{figure}

\section{\label{sec5} Custodial symmetry breaking}

    Adopting a term from the electroweak theory, we will define a ``custodial symmetry'' to be a global symmetry of one or more matter fields which (i) does not transform the gauge field; and for which (ii) any local operator which transforms non-trivially under the custodial symmetry also transforms
non-trivially under the local gauge symmetry.  The spontaneous or dynamical breaking of such a symmetry is therefore masked  by the unbroken gauge symmetry, which makes it difficult to see how to construct an order parameter for the custodial symmetry breaking without first fixing the gauge symmetry in some way.  We have already encountered one such symmetry, namely the transformation \rf{gtrans} with $\th(x)=\th$ independent of space.  Another symmetry of this kind is well known in the electroweak sector of the Standard Model.  Returning to the SU(2) lattice gauge-Higgs theory \rf{SU2gh}, we note that
the action is invariant under
\bea
            U_\m(x) &\ra& L(x)  U_\m(x) L^\dg(x+\hat{\mu})  \non \\
            \phi(x) &\ra& L(x) \phi(x) R \ ,
\eea
where $L(x) \in $ SU(2)${}_{gauge}$ is a local gauge transformation, while $R \in $ SU(2)${}_{global}$ is a global transformation.  SU(2)${}_{global}$ is sometimes referred to as the ``custodial'' symmetry of the theory, cf.\
 \cite{Maas:2019nso}.
                     
      It should be noted that if we choose a gauge (e.g.\ unitary gauge) in which the Higgs field acquires a vacuum expectation value
\beq
    \langle \phi \rangle = \left[ \begin{array}{cc}
                                               \upsilon & 0 \cr
                                               0 & \upsilon \end{array} \right] \ ,
\eeq      
then the SU(2)${}_{gauge} \times$ SU(2)${}_{global}$ symmetry is broken down to a diagonal global subgroup
\beq
\mbox{SU(2)}_{gauge} \times \mbox{SU(2)}_{global} \ra \mbox{SU(2)}_D \ ,
\eeq
corresponding to transformations
\bea
           L(x) &=& R^\dg =G  \non \\
           \phi(x) &\ra& G \phi(x) G^\dg  ~~,~~ U_\m(x) \ra G U_\m(x) G^\dg \ .
\eea
Some authors refer to transformations in this diagonal subgroup, which preserve the vacuum expectation value of
$\phi$ in a fixed gauge, as the custodial symmetry group. Whatever the terminology,
custodial symmetry has a role to play in the phenomenology of the electroweak interactions, and 
is reviewed in many places, e.g.\ \cite{Willenbrock:2004hu,Weinberg:1996kr,Maas:2019nso}.   Here, however, we wish
to focus first on the SU(2)${}_{global}$ group of $R$ transformations in the absence of gauge fixing, moving from there
to the $\th(x)=\th$ global U(1) symmetry group in the abelian theory.
  
      Does it make any sense to describe the Higgs phase of the theory as a phase of spontaneously broken
SU(2)$_{global}$ symmetry, what we call here custodial symmetry?   Local gauge symmetries cannot break according to
the Elitzur theorem, and the breaking of a global subgroup of the gauge symmetry appears to depend on the gauge choice, as we have seen in the previous section.  There is also no gauge-invariant local order parameter for custodial symmetry breaking, so it cannot break spontaneously in the usual sense (and if it did, one would have to contend with the Goldstone theorem).  On the other hand, the full partition function of the SU(2) gauge-Higgs theory can regarded
as a sum of partition functions of a spin system in an external gauge field, i.e.    
\beq
               Z  =  \int DU ~ Z_{spin}[U] e^{-S_W(U)} \ ,
\eeq
where
\beq
               Z_{spin}[U] = \int D\phi ~ e^{-S_H(U,\phi)} \ ,
\eeq
and, depending on $U$, custodial symmetry {\it can} break in the system described by $Z_{spin}(U)$.  

   Let us define the expectation value of an operator $\Omega[U,\phi]$ in the spin system
\beq
       \overline{\Omega}(U) = {1\over Z_{spin}(U)} \int D\phi \Omega(\phi,U) e^{-S_H} \ ,
\eeq
with the full expectation value
\bea
          \langle \Omega \rangle &=& \int DU P(U) \overline{\Omega}(U) \non \\
                                               &=& {1\over Z} \int DU D\phi \Omega(\phi,U) e^{-S} \ .
\eea
This means that the expectation value in the spin system is to be evaluated from ensembles with $U$ chosen
from the probability distribution
\beq
          P(U) = {1\over Z} Z_{spin}[U] e^{-S_W(U)} \ .
\label{PU}
\eeq
So the question becomes: is $Z_{spin}[U]$ in the broken or the unbroken phase, for gauge field configurations selected from this probability distribution?   It is not hard
to devise a gauge-invariant operator $\Phi(U)$ which is non-zero in the broken phase, and which vanishes in the unbroken
phase in the thermodynamics limit.   Then $\langle \Phi \rangle \ne 0$, i.e.\ custodial symmetry breaking, is our proposed
{\it definition} of the Higgs phase of a gauge-Higgs theory. 

     In a numerical simulation we may determine whether $Z_{spin}(U)$ is in the broken phase in the probability distribution
defined by \rf{PU} by a 
``Monte Carlo-within-a-Monte Carlo simulation.''  The procedure is to update the $U_\m(x), \phi(x)$ fields in  the full gauge-Higgs theory in the usual way
for, e.g., 100 update sweeps, which is followed by the data-taking procedure, which is itself a lattice Monte-Carlo simulation of $Z_{spin}(U)$, keeping the link variables fixed at whatever they were at the end of the last update sweep.  
The $Z_{spin}(U)$ simulation proceeds for  $n_{spin}$ sweeps, updating only the $\phi(x)$ variables.  Let $\phi(x,n)$ denote $\phi(x)$ at the $n$-th update sweep of the spin system, and let 
\beq
           \overline{\phi}_{n_{spin}}(x) = {1\over n_{spin}}\sum_{n=1}^{n_{spin}} \phi(x,n)  \ .
\eeq
We then define
\beq
 \Phi_{n_{spin},V}[U] = {1\over V} \sum_x | \overline{\phi}_{n_{spin}}(x) |  \ ,
\eeq
where $|\phi| = \det^{\oh}(\phi)$, and
\beq
           \Phi[U] = \lim_{n_{spin} \ra \infty} \lim_{V \ra \infty} \Phi_{n_{spin},V}[U]  \ .
\eeq
Averaging $\Phi_{n_{spin},V}[U]$ over many data-taking sweeps at large $n_{spin}$, and extrapolating to infinite volume,
provides a numerical estimate of $\langle \Phi[U] \rangle$.
 Then the Higgs phase of the full gauge-Higgs theory is distinguished from the unbroken phase by
\beq
            \langle \Phi[U] \rangle = \left\{ \begin{array} {cl} 
                                             \text{zero} & \text{unbroken phase} \cr
                                             \text{non-zero} & \text{Higgs phase} \end{array} \right. \ .
\eeq
This procedure was carried out for the SU(2) and SU(3)
gauge-Higgs models in ref.\  \cite{Greensite:2018mhh}, where we have determined the transition line between 
the phases of broken and unbroken custodial symmetry, as defined above.  The custodial symmetry breaking
transition in the SU(2) theory is shown in Fig.\ \ref{su2phase}, together with the remnant symmetry breaking line
for Landau gauge. At the larger $\b$ values the two transitions coincide, and also coincide with a sharp crossover
in the action vs.\ $\g$, which is also shown.  The Coulomb transition line (not shown, but see Fig.\ \ref{su2_break}), lies above the Landau transition. 

\begin{figure}[h!]   
\centerline{\includegraphics[scale=0.6]{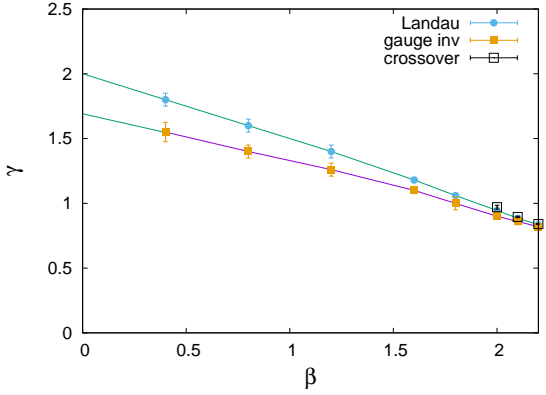}}
\caption{The custodial symmetry (labeled ``gauge inv'' ) and Landau gauge remnant symmetry transition
points in SU(2) gauge-Higgs theory.  Points labeled ``crossover'' locate a sharp thermodynamic crossover, but
not a phase transition.  Note that the Landau transition lies above the line of custodial symmetry breaking.}
\label{su2phase}
\end{figure}    

    We may define the gauge-invariant observable $\Phi[U]$ more formally, without any appeal to numerical simulations, by introducing a small perturbation which is removed after taking the thermodynamic limit.  Let
\bea
            \overline{\phi}_{JV}[x;U,\eta] &=& {1 \over Z_{spin}[U]} \int D\phi \phi(x) \non \\
                       &\times& \exp\left[ - S_H + J \sum_x \tr[\eta^\dg(x) \phi(x)] \right]  \ ,
\eea
where $\eta(x)$ is a unimodular field $|\eta|=1$, which is chosen to be any one of an equivalent set of configurations,
related by the SU(2)$_{global}$ symmetry, which maximizes the averaged sum of moduli
\beq
            \Phi_{JV}[U] = \max_{\eta} {1\over V} \sum_x | \overline{\phi}_{JV}[x;U,\eta] |  \ .
\label{eta}
\eeq
We then define the order parameter for symmetry breaking
\beq
           \langle \Phi \rangle = \lim_{J\ra 0} \lim_{V \ra \infty} \langle \Phi_{JV}[U] \rangle \ ,
\eeq
with the order of limits as shown.  This parameter is non-zero if the SU(2)$_{global}$ symmetry of the spin system is spontaneously broken, and zero otherwise.   We observe that $\Phi[U]$ is manifestly gauge invariant, with a vacuum expectation value determined in the full gauge-Higgs theory.  

The field $\eta(x)$ which maximizes the right hand side of \rf{eta} for a given $U_\m(x)$ configuration is very difficult to determine in practice.  Since no gauge is fixed, $U_\m(x)$ varies wildly in space, and the same will be true of $\eta(x)$.  Were we to define the spatial average of $\overline{\phi}_J[x;U,\eta]$ before taking the modulus, it would average to zero in general.  In practice we use the lattice Monte Carlo procedure, described
above, to determine $\langle \Phi \rangle$.

\subsection{Significance}

     As we have already emphasized, spontaneous symmetry breaking can only occur for a global subgroup of the
 gauge group, and only in a gauge which leaves unfixed a remnant global symmetry of the gauge group. We are interested
 in gauges for which $\langle \phi \rangle \ne 0$ necessarily implies the spontaneous breaking of some global subgroup of the gauge group. Unitary gauge is excluded by this restriction, since in that gauge $\langle \phi \rangle \ne 0$ in any phase, including the massless phase, independent of the dynamics.   Let us consider instead gauge conditions $F(U)=0$. At a minimum, such a gauge condition leaves unfixed a global transformation $g(x) = z$, where $z$ is an element of the
center of the gauge group.   Of course, gauge conditions of this kind may have a larger remnant symmetry, e.g.\ Landau gauge has a remnant symmetry $g(x)=R$ where $R$ is any element of the SU(2)$_\text{global}$ as discussed above, but in any case the global subgroup of the gauge group consisting only of center elements is always a remnant symmetry in gauges of this kind.  In the case of U(1) symmetry, the center subgroup is the group itself.  It is important to note here that in general the global center transformations belong both to the gauge group, and to the custodial symmetry group as defined above.
 
    We now make the following observations:
     
\begin{enumerate}
\item Custodial symmetry breaking is a necessary condition for the spontaneous breaking of a global subgroup of the gauge group in any given gauge. 
\item Custodial symmetry breaking is a sufficient condition for the existence of {\it some} gauge in which a global subgroup of the gauge group is spontaneously broken.  
\end{enumerate}

    Start with the first point.  Stated a little more precisely, consider any gauge condition $F(U)=0$ which leaves unfixed
a global subgroup of the gauge symmetry, and let
\bea
    | \langle \phi \rangle_{JV}| &=& {1\over Z} \left| \int DU D\phi  \D[U] \d[F(U)] \left({1\over V} \sum_x \phi(x) \right) 
    e^{-S} \right. \non \\
     & & \qquad \qquad \times \left. \exp[J\sum_x \tr(\phi(x))] \right|
\label{phiJV}
\eea
in volume $V$ where $\D[U]$ is the Faddeev-Popov term.  The global subgroup of the gauge symmetry is said to be spontaneously broken in this gauge if
\beq
    |\langle \phi \rangle| = \lim_{J\ra 0} \lim_{V\ra \infty} |\langle \phi \rangle_{JV}| \ne 0 \ .
\eeq
The statement is that symmetry breaking of that kind is only possible if $\langle \Phi \rangle > 0$, i.e.\ if constituent
symmetry is also spontaneously broken.  This can be seen from the definition of constituent symmetry breaking.  Since
$\Phi_{JV}(U)$ is gauge invariant, it can of course be evaluated with or without gauge fixing, and in particular in
the gauge $F(U)=0$.  Then
\begin{widetext}
\bea
     \langle \Phi_{JV} \rangle &=& {1\over Z} \int DU \D[U] \d[F(U)] e^ {-S_W}  Z_{spin}[U]  \max_{\eta}
      {1\over V} \sum_x | \overline{\phi}_J[x;U,\eta] | \non \\
   &=& {1\over Z} \int DU \D[U] \d[F(U)] e^ {-S_W} Z_{spin}[U]  \Big\{ {1\over Z_{spin}[U]} \max_{\eta}  {1\over V}\sum_x
       \left| \int D\phi \phi(x) e^{-S_H} \exp[ J \sum_x \tr[\eta^\dg(x) \phi(x)] ] \right| \Big\} \non \\
   &=& {1\over Z} \int DU \D[U] \d[F(U)] e^ {-S_W} \max_{\eta}  {1\over V}\sum_x \left| \int D\phi \phi(x)  e^{-S_H}
           \exp[ J \sum_x \tr[\eta^\dg(x) \phi(x)] ] \right|  \ .
\eea
However, from \rf{phiJV}
\bea
 | \langle \phi \rangle_{JV}| &\le& {1\over Z} \int DU \D[U] \d[F(U)] e^ {-S_W} \left| {1\over V} \sum_x
      \int D\phi \phi(x) e^{-S_H}  \exp[J\sum_x \tr(\phi(x))] \right|
\non \\
&\le& {1\over Z} \int DU \D[U] \d[F(U)] e^ {-S_W} {1\over V} \sum_x
      \left| \int D\phi \phi(x) e^{-S_H}  \exp[J\sum_x \tr(\phi(x))] \right|
\non \\
&\le& {1\over Z} \int DU \D[U] \d[F(U)] e^ {-S_W} \max_{\eta} {1\over V} \sum_x
     \left| \int D\phi \phi(x) e^{-S_H}  \exp[J\sum_x \tr(\eta^\dg(x) \phi(x))] \right|
\non \\ 
&\le& \langle \Phi_{JV} \rangle \ .
\eea
\end{widetext}
Taking first the infinite volume and then the $J\ra 0$ limits, it follows that
\beq
           \langle \Phi \rangle \ge |\langle \phi \rangle|  \ .
\eeq
So although remnant gauge symmetry may or may not be broken at some point in the space of couplings, depending on
the choice of gauge, we can conclude that the existence of spontaneous gauge symmetry breaking for those couplings in {\it some} gauge is only possible if custodial symmetry is also spontaneously broken.   This means, in particular, that 
the custodial symmetry breaking line must lie below the remnant gauge symmetry breaking lines in Coulomb and Landau
gauges, which is indeed what we see in Fig.\ \ref{su2phase}, taken together with Fig.\ \ref{su2_break}.

   Moving on to the second point, let us define $\overline{\phi}_{JV}(x;U)$ as $\overline{\phi}_{JV}(x;U,\eta)$ with $\eta$
chosen to maximize the right hand side of \rf{eta}.  Let
\beq
\widehat{\phi}_{JV}(x;U) = {\overline{\phi}_{JV}(x;U) \over |\overline{\phi}_{JV}(x;U)| }
\eeq
and we consider the gauge
\beq
        \widehat{\phi}_{JV}(x;U) = \mathbbm{1}
\eeq
Since this condition is imposed only on the gauge field, there is obviously a remnant gauge symmetry under those
transformations which leave $U$ invariant.  For the SU(N) gauge-Higgs theories this is a global center symmetry, while
in the $q=2$ abelian Higgs model it is the global transformations under U(1)/$Z_2$.
In this special gauge,  introducing an explicit breaking term
\bea
& & | \langle \phi \rangle |  = 
 \lim_{J \ra 0} \lim_{V \ra \infty} {1\over Z} \bigg| \int DU \D[U] 
           \d[\widehat{\phi}_{JV}(x;U)-\mathbbm{1}] e^{-S_W} \non \\
         & & \qquad \times \max_{\eta}  \int D\phi  {1\over V} \sum_x \phi(x)  e^{-S_H}  
         \exp[J \sum_x \tr(\eta^\dg(x) \phi(x))]  \bigg| \non \\
 & & \qquad = \lim_{J \ra 0} \lim_{V \ra \infty} {1\over Z} \int DU \D[U] 
           \d[\widehat{\phi}_{JV}(x;U)-\mathbbm{1}]  e^{-S_W} \non \\
         & & \qquad \times \max_{\eta}  {1\over V} \sum_x \left| \int D\phi \phi(x)  e^{-S_H}  
         \exp[J \sum_x \tr(\eta^\dg(x) \phi(x))] \right|
\non \\
         & & \qquad  = \langle \Phi \rangle
\eea     
If custodial symmetry is spontaneously broken, then ${|\langle \phi \rangle| > 0}$, and the remnant global gauge symmetry is also spontaneously broken.

    A custodial symmetry in a non-abelian theory is not necessarily a continuous symmetry.  Let us consider a
lattice version of an SU(N) gauge-Higgs theory, this time with the unimodular Higgs field in the adjoint representation.  A lattice action with the correct continuum limit is \cite{Drouffe:1984hb}
\beq
     S =  -S_W  - \g \sum_{x,\m} \tr[\Gamma(x) U_\m(x) \Gamma^\dg(x+\hat{\m}) U^\dg_\m(x)] \ ,
\eeq
where $\Gamma(x)$ is an SU(N)-valued Higgs field, and $S_W$ is the usual Wilson action.  The custodial symmetry in this case is the discrete global symmetry
\beq
            \Gamma(x) \ra \Gamma'(x) = z_n \Gamma(x) \ ,
\eeq
where
\beq
  z_n =e^{2\pi i n/N},~n=0,1,...,N-1 \in Z_N \ ,
\eeq
and the set of elements $\{ z_n \mathbbm{1} \}$ constitute the center subgroup of SU(N).

\subsection{Custodial symmetry breaking in the abelian Higgs model}

    After this excursion into non-abelian gauge theory we return to the example relevant to superconductivity, i.e.\ the lattice abelian Higgs model \rf{ahiggs} with a double-charged Higgs field, corresponding to $q=2$.   We observe that
the action is invariant under a global U(1) transformation $\phi(x) \ra e^{i\a} \phi(x)$ of the Higgs field alone.
By our definition this is a custodial symmetry, in this case indistinguishable from a global gauge transformation, whose spontaneous breaking can be detected by the methods outlined above.   In numerical
simulations we use the Monte-Carlo-within-a-Monte-Carlo approach, calculating $\Phi_{n_{spin},V}[U]$ during the
data taking process using $n_{spin}$ update sweeps of the $\phi$ field at fixed $U$, and averaging over the values obtained at every set of data-taking sweeps at fixed $U$ to arrive at $\langle \Phi_{n_{spin},V}[U] \rangle$.  This quantity is computed at a range of $n_{spin}$ on a $V=12^4$ lattice volume, and extrapolated to $n_{spin}=\infty$ by fitting the data to
\beq
          \langle \Phi_{n_{spin},V}[U] \rangle = \langle \Phi_V[U] \rangle +     {\mbox{const.} \over  \sqrt{n_{spin}}}   \ .    
\eeq
Below the transition line, $\langle \Phi_V[U] \rangle=0$, while above the line $\langle \Phi_V[U] \rangle > 0$.
An example of this procedure is shown in Fig.\ \ref{b05}, where we present data for 
$\langle \Phi_{n_{spin},V}[U] \rangle$ vs.\ $n_{spin}$  at $\b=0.5$, at $\g$ values above ($\g=0.9$) and below
($\g=0.7$) the transition.  

\begin{figure}[h!]   
\centerline{\includegraphics[scale=0.7]{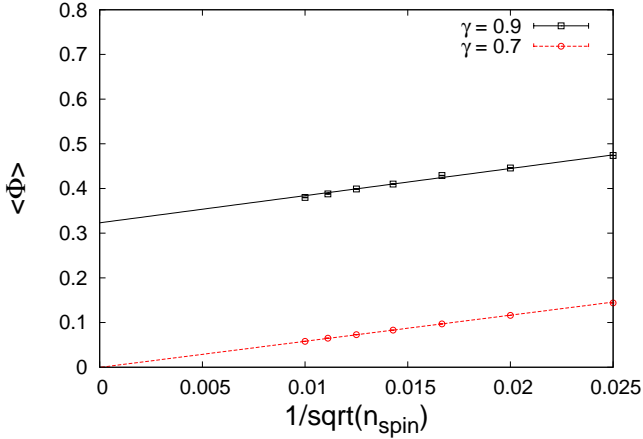}}
\caption{The parameter $\langle \Phi_{n_{spin}} \rangle$ vs. $1/\sqrt{n_{spin}}$ at $\b=0.5$.  The custodial symmetry
breaking transition is at $\g_c \approx 0.84$.  The plot displays our values of $\langle \Phi \rangle$ below ($\g=0.7$)
and above ($\g=0.9$) the critical value.  For all $\g < \g_c$, $\langle \Phi \rangle$ extrapolates to zero as
$n_{spin} \ra \infty$, while at all $\g > \g_c$ $\langle \Phi \rangle$ extrapolates to a non-zero value.}
\label{b05}
\end{figure}

\begin{figure}[h!]   
\centerline{\includegraphics[scale=0.7]{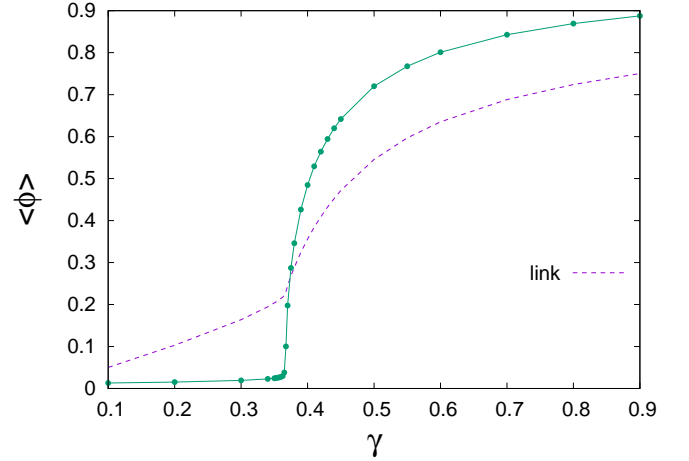}}
\caption{Order parameter $\langle \Phi \rangle$ for custodial symmetry breaking vs.\ $\g$ at $\b=2.0$ on
a $16^4$ lattice volume, with $n_{spin}=6400$.  Also shown (dashed line) is the corresponding data for $L$ vs.\ $\g$,
already shown in Fig.\ \ref{kink}.  The thermodynamic and custodial symmetry breaking transitions coincide at
$\g=0.365$.}
\label{cust20}
\end{figure}

At points where the custodial symmetry transition coincides with the thermodynamic transition, there is an
abrupt rise in $\langle \Phi_{n_{spin}} \rangle$ even at moderate values of $n_{spin}$ as illustrated in Fig.\ \ref{cust20},
where we plot $\langle \Phi \rangle$ vs.\ $\g$ at $\b=2$ on a $16^4$ lattice.  Also shown in this figure, as a dashed line
is the corresponding data for the average link variable $L$, already displayed in Fig.\ \ref{kink}.  It is clear that the
thermodynamic transition (the ``kink'') and custodial breaking transition, signalled by a sudden rise in $\langle \Phi \rangle$, occur at the same point, namely $\g=0.365$ at $\b=2$.

The custodial symmetry transition line in the $\b-\g$ plane is shown in Fig.\ \ref{cust}.

\begin{figure}[h!] 
\centerline{\includegraphics[scale=0.7]{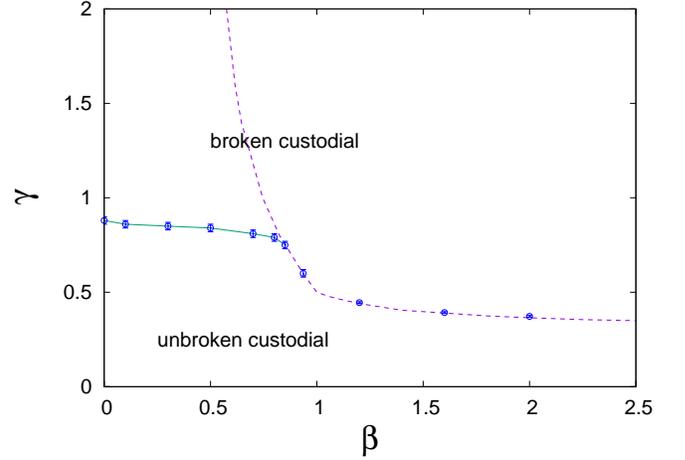}}
\caption{Custodial symmetry breaking transition points.  The dashed line is a line of thermodynamic transition shown
in Fig.\ \ref{thermo}, which coincides with custodial symmetry breaking at $\b > 0.85$.  For lower values of $\b$, the
custodial symmetry breaking transition occurs in the region labeled ``conf'' in Fig.\ \ref{thermo}.}
\label{cust}
\end{figure}   
    
\subsection{Absence of Goldstone Excitations}

    The reason that spontaneous breaking of custodial symmetry does not result in physical gapless excitations is
essentially the same reason given long ago \cite{Guralnik:1967zz,*Guralnik:1964eu}, when a similar question was raised regarding the spontaneous breaking of (remnant) gauge symmetries.  In the case of an abelian theory it is 
obvious that the same reasoning must apply, because in that case the custodial symmetry is identical to the remnant gauge symmetry $\th(x) = \th$.

    In a little more detail, spontaneous breaking of custodial symmetry in a given $Z_{spin}(U)$ for some $U$ may very well be associated with gapless excitations.  However, there is no reason to believe that such excitations appear in correlation functions associated with physical states.  For instance, if custodial symmetry is broken in $Z_{spin}(U)$, with order parameter $\phi(x)$, then for fixed $U$ there might be a long-range part to a correlator such as
\beq
          \overline{\phi(x)\phi(y)} - \overline{\phi(x)} \times \overline{\phi(y)}
\eeq
Such a correlator however, being locally gauge non-invariant, would necessarily vanish in the full theory, i.e.
\beq
          \langle \phi(x)\phi(y) \rangle - \langle \phi(x) \rangle \langle \phi(y) \rangle = 0
\eeq
In order to apply the Goldstone theorem to custodial symmetry, and restrict to physical excitations,  it is necessary to fix to a gauge which eliminates extraneous degrees of freedom, leaving only physical degrees of freedom.  Examples are Coulomb gauge and axial gauge. In such gauges it is necessary to impose Gauss's Law as an operator identity, and solve for $E_L^2$ in the Hamiltonian.  Gauges of this type are Lorentz non-invariant, and the $E_L^2$ term gives rise to long-range interactions in the Hamiltonian.   Non-local terms in general violate one of the assumptions of the Goldstone theorem.   This observation was made originally
in reference to the breaking of remnant gauge symmetries \cite{Guralnik:1967zz,*Guralnik:1964eu}, but it applies equally
well to the current associated with any continuous custodial symmetry.  The conclusion is that spontaneous breaking of a global custodial symmetry does not necessarily imply gapless physical excitations, which might have been expected from the Goldstone theorem.

\subsection{\label{CSc} C vs S$_\text{c}$ Confinement}

   Custodial symmetry, and also remnant gauge symmetry in Coulomb and Landau gauges, have transition lines
in the confinement region of the $q=2$ phase diagram.   Usually $\langle \phi \rangle \ne 0$ is associated with a Higgs phase, so how can this happen in a confined phase?  In this case it is helpful to consider unitary gauge at large $\g$,
and write the link variables in the form $U_\m(x) = \tU_\m (x) Z_\m(x)$, where Re$[\tU_\m(x)]>0$ and 
$Z_\m(x) = \pm 1$.  As $\g \ra\infty$, then $\tU_\m(x) \ra 1$, and the abelian Higgs model goes over to $Z_2$ lattice gauge
theory, which has a confined and unconfined phase. But what is confined, in the confined phase, are $q=1$ test charges,
 i.e.\ sources with $\pm 1$ units of electric charge. Test charges with $q=2$ are insensitive to the $Z_\m(x)$ degrees of freedom, and couple only to $\tU_\m(x)$.
Away from unitary gauge, the remnant gauge symmetry which is broken spontaneously by $\langle \phi \rangle \ne 0$ is global $U(1)/Z_2$, and from the point of view of $q=2$ sources the theory is actually in a Higgs phase.  This raises the question of the nature of the transition, as seen by $q=2$ sources, from the confined phase into the Higgs phase, since by criteria such as Wilson loops and Polyakov lines the $q=2$ sources are not really confined anywhere in the phase diagram.

     We have addressed the same question in ref.\  \cite{Greensite:2018mhh}, in the context of SU(2) gauge-Higgs theory
with the Higgs field in the fundamental representation of the gauge group.  In this theory, as in any gauge theory with matter in the fundamental representation (such as QCD), Wilson loops fall off asymptotically with a perimeter law, and Polyakov lines have a non-zero vacuum expectation value.  Then what is meant by the word ``confinement''
in such theories?  A common answer is that confinement means that only color singlet particle states appear in the
asymptotic spectrum, a property which we will refer to as ``C-confinement.''   It is well known that this property holds not
only in confinement-like region of an SU(2) gauge-Higgs theory, but also deep in the Higgs regime 
\cite{Osterwalder:1977pc,Fradkin:1978dv,Frohlich:1981yi,tHooft:1979yoe}.  Nevertheless there seems to be a qualitative difference between these regions, since in the confinement-like region there is color electric flux tube formation, linear Regge trajectories, and a linear potential up to string breaking, as in QCD,  while in the Higgs region there is no electric flux tube formation in any distance regime, no linear Regge trajectories, and only Yukawa forces among particles.

   In a pure SU(2) gauge theory, the word ``confinement''  includes but goes beyond the property of C confinement.
Certainly the asymptotic spectrum consists only of color singlets, i.e.\ glueballs.  But it also has the property that
the energy $E(R)$ above the vacuum energy, of any physical state containing a static quark-antiquark pair, is bounded from below by a linear potential.  In other words, let $V_{ab}(x,y;A)$ be any functional of the gauge field $A$ which transforms covariantly under the gauge group, and we consider physical states of the form
\beq
            \Psi_V = \overline{q}^a(\bx) V_{ab}(\bx,\by;A) q^b(\by) \Psi_0
\eeq
where $\Psi_0$ is the ground state.  Let $E_V(R) = \langle \Psi_V | H-E_0 | \Psi_V \rangle$ be the expectation value of energy, above the vacuum energy $E_0$, in state $\Psi_V$, where $R=|\vx-\vy|$.  We define ``separation-of-charge'' confinement, or ``S$_\text{c}$'' confinement for short,
to mean that $E_V(R)$ is bounded from below, asymptotically, by a linear potential
\beq
          \lim_{R \ra \infty} {dE_V \over dR} > \s
\label{Ev}
\eeq
for {\it any} choice of $V$.  Pure SU(N) gauge theories in $D \le 4$ dimensions certainly have this property.  We have
suggested in \cite{Greensite:2018mhh} that this same definition extends to gauge theories with matter fields, with the
essential requirement that $V(x,y;A)$ depends only on the gauge field, and not on the matter fields.  This restriction
essentially tests whether the dynamics would form a flux tube between sources if we exclude string breaking by
matter fields. In the cited reference we have shown that there must exist a transition between the C and S$_\text{c}$ confinement regions, and we have also computed, in SU(2) and SU(3) gauge-Higgs theories, the line of custodial symmetry breaking.  Our conjecture, for which we have presented some evidence, is that the S$_\text{c}$-to-C confinement transition, and the custodial symmetry breaking transition, coincide. 

   That is also our conjecture regarding the custodial symmetry breaking transition inside the confinement phase of the $q=2$ abelian Higgs model, with this modification:  For the $q=2$ theory we have confinement of single charged ($q=1$) sources
by a linear potential whenever the $Z_2$ global symmetry defined in section \ref{z2sec} is unbroken, which is the entire region labeled ``conf'' in Fig.\ \ref{thermo}.   The C-vs-S$_\text{c}$
transition in the $q=2$ theory concerns the nature of the confined phase for {\it double}-charged ($q=2$) objects, which are insensitive to the $Z_2$ degrees of freedom.  Double-charged Wilson loops have a perimeter law falloff and double-charged Polyakov lines are non-zero inside the confined phase, as in SU(N) gauge theories (such as QCD) with matter in the fundamental representation.  We can define S$_{c}$ confinement for 
double charged sources in the same way:  Consider operators $V(\bx,\by;A)$, and $q=2$ matter fields $\psi(x)$ which transform under a gauge transformation $g(x) = \exp[i\th(x)]$ as
\bea
          V(\bx,\by;A) &\ra&    e^{2i\th(\bx)} V(\bx,\by;A) e^{-2i\th(\bx)} \non \\
          \psi(x) &\ra& e^{2i\th(\bx)} \psi(x) \non \\
          \Psi_V &=&  \overline{\psi}(\bx) V(\bx,\by;A) \psi(\by) \Psi_0
\eea
Then the theory is S$_\text{c}$ confining when the condition
\rf{Ev} is satisfied.  As in the non-abelian theory, our conjecture is that custodial symmetry breaking at small $\b$ coincides
with the transition from S$_\text{c}$ to C  confinement for $q=2$ charges.   

      Our point is this:  from the standpoint of $q=2$ charged matter in a $q=2$ abelian Higgs theory, the transition from a confined phase (which we define as S$_\text{c}$ confinement) to a Higgs phase need not coincide everywhere with the transition from a confined to a Higgs phase for $q=1$ test charges.  What we are proposing is that the spontaneous breaking of
custodial symmetry is a gauge invariant criterion which sets the boundary of the Higgs region, as seen by $q=2$ matter in the $q=2$ abelian Higgs theory. 
 
    We should finally note that the custodial and remnant gauge symmetry breaking lines in the confinement region of the $q=2$ gauge-Higgs model, and also in the SU(2) gauge-Higgs theory, are not lines of thermodynamic transition.  As we have just argued, this does not imply irrelevance. Recall that there are other physically meaningful transitions in statistical systems which, like custodial and remnant symmetry breaking, are not necessarily associated with thermodynamic transitions.  We here have in mind the geometric transition lines, also known as Kertesz lines, in Ising and Potts models, which are associated with percolation transitions \cite{Kertesz,Blanchard_2008} .

\section{\label{conclude} Conclusions}
         
      In this article we have pointed out that  ``spontaneous breaking of gauge symmetry'' is an ambiguous concept, and
we have proposed that it is spontaneous breaking of custodial symmetry which characterizes the Higgs phase.  The ambiguity of spontaneous gauge symmetry breaking is due to the fact that local gauge symmetries cannot break spontaneously, as we know from the Elitzur theorem, which means that only a global subgroup of the gauge symmetry can break spontaneously, and this is visible only in a gauge which leaves this global subgroup unfixed.  This means that the order parameter for spontaneous gauge symmetry breaking is gauge dependent.  As shown previously for SU(2) gauge-Higgs theory, and as shown here in the lattice abelian Higgs model, spontaneous gauge symmetry breaking can occur at different places in the phase diagram in different gauges, and in some gauges it may even disappear entirely.  We might add that in unitary gauge, in  U(1) and SU(2) gauge-Higgs theories,  there is no global gauge symmetry which remains that can break spontaneously. In this gauge the scalar field has a non-zero expectation value in any phase, Higgs or massless or confining, just due to quantum fluctuations (or, in some theories, due to the fact that the scalar field is taken to have a fixed modulus from the beginning). 

     Adopting a term from the electroweak theory, we have defined ``custodial symmetry'' to be (i) a group of transformations
of the matter fields which does not transform the gauge field, and (ii) a symmetry for which there is no gauge invariant
order parameter, in the sense that any operator which transforms under the custodial symmetry also transforms under the gauge group.   The custodial symmetry group and global gauge transformations share symmetry transformations
which belong to the center of the gauge group, which for U(1) gauge theory is the group itself.  Despite the absence of a gauge invariant order parameter, we have shown here how spontaneous breaking of the custodial symmetry can be defined and observed in a gauge-invariant manner, without recourse to gauge fixing.

    The relation of custodial symmetry breaking to gauge symmetry breaking is as follows:  First, custodial symmetry breaking is a necessary condition for gauge symmetry breaking in any particular gauge.  Secondly, custodial symmetry
is a sufficient condition for the existence of some gauge in which the gauge symmetry breaks spontaneously.
If we identify the Anderson-Brout-Englert-Higgs mechanism with the existence of spontaneous gauge symmetry breaking in some gauge, then this mechanism occurs if and only if custodial symmetry is spontaneously broken.  In the $q=2$ abelian Higgs model, we have seen numerically that custodial symmetry breaks along the line separating the massless
and Higgs phases.
 
     In some regions of the phase diagram, gauge symmetries and custodial symmetry can break without a corresponding thermodynamic transition, as is the case for the geometric (Kertesz) transition in the Ising and Potts models.  We believe
that custodial symmetry breaking in the absence of a thermodynamic transition is related to what we have elsewhere
described as the transition between separation-of-charge confinement and color confinement \cite{Greensite:2018mhh}.  This correspondence is so far a conjecture, and calls for further investigation.
       
\appendix*

\section{}

  Here we re-derive the fact that the photon remains massless in the normal phase, which requires an exact
cancellation at $k=0$ between the terms inside the square root in eq.\ \rf{muq}.  We assume that the energy gap
$\D$ vanishes, and all energy levels are filled up to the Fermi surface.  Then
\bea
r(q) |A^T(q) |^2 &=& {1\over V} {e^2 \over m^2} \sum_p  {(u_p v_{p-q} - u_{p-q} v_p)^2 \over 
             \left| {p^2 \over 2m} - \e_F \right| +  \left| {(p-q)^2 \over 2m} - \e_F \right| } \non \\
             & & \times |\vp \cdot  A^T(q) |^2  \ .
\label{r1}
\eea
In order to compute $r(q)$ at small $q$, we let $\vq$ define the positive $z$-direction, and then $A^T$ is perpendicular
to $z$, and we can take the polarization to lie  along, e.g., the $x$-direction.  Approximating the mode sum by an
integral over continuous wavenumbers, and cancelling $|A^T(q) |^2$ on both sides, we have
\bea
r(q)  &=& {1\over (2\pi)^3} {e^2 \over m^2} \int d^3p {(u_p v_{p-q} - u_{p-q} v_p)^2 p_x^2 \over
                                  \left| {p^2 \over 2m} - \e_F \right| +  \left| {(p-q)^2 \over 2m} - \e_F \right| } \ .
\label{r2}
\eea
If $v=1,u=0$ below the Fermi surface, and $v=0,u=1$ above, then $(u_p v_{p-q} - u_{p-q} v_p)^2 = 1$
if $\vp$ and $\vp-\vq$ lie on opposite sides of the Fermi surface, and equals zero otherwise.  Let $p_F = \sqrt{2m \e_F}$.  In the ``northern'' hemisphere, i.e.\ $p_z>0$, and small $q\ll p_F$,  the integration region will be consist of momenta $\vp$ outside the Fermi surface, and $\vp-\vq$ inside; and vice-versa in the southern hemisphere.  Both
hemispheres give the same contribution, so it will be enough to compute the contribution in the northern hemisphere
and multiply by two.   Let 
\beq
             \vp = (p_F + k) \hat{e}_p  \ ,
\eeq
where $\hat{e}_p$ is a unit vector in the $\vp$ direction.  Then we require
\bea
          p_F^2 &>& |\vp - \vq|^2 \non \\
                     &>& p_F^2  + 2p_F k - 2p_F  q \cos(\th) + O(q^2) \ ,
\eea
where $\th$ is the angle to the $p_z$ axis.  Dropping the O($q^2$) term at small $q$, the condition is
\beq
           0 < k < q \cos(\th) \ .
\eeq
Also dropping the O($q^2$) term in the denominator of \rf{r2}, we have in this region
\beq
\left| {p^2 \over 2m} - \e_F \right| +  \left| {(p-q)^2 \over 2m} - \e_F \right| \approx {q p_z \over m} \ .
\eeq
Then we have, for $q \ra 0$
\bea
    r(q) &=& {1 \over (2\pi)^3} {e^2 \over m q} 2 \int_{\cos(\th)>1} D\Omega \int_{p_F}^{p_F+q \cos(\th)} dp p^2 
    {p_x^2 \over p_z} \non \\
           &=& {1 \over (2\pi)^3}  {4 \pi \over 3} {e^2 \over m} p_F^3 \ ,
\eea
and using the relation
\beq
           p_F^3 = {3 \over 8\pi} (2\pi)^3 \eta \ ,
\eeq
we find that
\beq
           r(q) = {e^2 \over 2m} \eta \ ,
\eeq
at $q\ra 0$, and the photon mass $\m(q=0)$ is zero for the gapless state, as it should be.

\acknowledgments{We thank Aron Beekman and Eduardo Fradkin for helpful correspondence. This work is supported by the U.S.\ Department of Energy under Grant No.\ DE-SC0013682.}  

\bibliography{bcs}

\end{document}